\documentclass[fleqn,usenatbib]{mnras}

\usepackage{newtxtext,newtxmath}
\usepackage[T1]{fontenc}

\DeclareRobustCommand{\VAN}[3]{#2}
\let\VANthebibliography\thebibliography
\def\thebibliography{\DeclareRobustCommand{\VAN}[3]{##3}\VANthebibliography}

\usepackage{graphicx}	
\usepackage{amsmath}	
\usepackage{tablefootnote}
\usepackage{stfloats}
\usepackage{subcaption}
\usepackage{multirow}

\usepackage{color}
\definecolor{bbsalmon}{rgb}{1.0, 0.47, 0.42}
\definecolor{bbgreen}{rgb}{0.01, 0.58, 0.53}

\usepackage{xcolor}

\title[Evidence for SMBH recoils in AGNs]{Statistical evidence for massive black hole recoils in active galactic nuclei}

\author[B.~B\'ecsy et al.]{
Bence B\'ecsy,$^{1}$\thanks{E-mail: b.becsy@bham.ac.uk (BB)}
Peter Raffai,$^{2,3}$
Zolt\'an Haiman,$^{4,5,6}$
Andor Budai,$^{2}$
and Zsolt Frei$^{2,3,7}$
\\
$^{1}$Institute for Gravitational Wave Astronomy and School of Physics and Astronomy, University of Birmingham, Edgbaston, Birmingham B15 2TT, UK\\
$^{2}$Institute of Physics and Astronomy, ELTE E\"otv\"os Lor\'and University, 1117 Budapest, Hungary\\
$^{3}$HUN-REN–ELTE Extragalactic Astrophysics Research Group, 1117 Budapest, Hungary\\
$^{4}$Institute of Science and Technology Austria (ISTA), Am Campus 1, Klosterneuburg 3400, Austria\\
$^{5}$Department of Astronomy, Columbia University New York, NY 10027, USA\\
$^{6}$Department of Physics, Columbia University New York, NY 10027, USA
\\
$^{7}$Konkoly Observatory, HUN-REN Research Centre for Astronomy and Earth Sciences, H-1121 Budapest, Konkoly Th.M. 15-17, Hungary
}

\date{Accepted XXX. Received YYY; in original form ZZZ}

\pubyear{2026}

\begin{document}
\label{firstpage}
\pagerange{\pageref{firstpage}--\pageref{lastpage}}
\maketitle

\begin{abstract}
We search for a population-level signature of gravitational-wave recoiling supermassive black holes: a positive correlation between dust obscuration and the magnitude of the line-of-sight velocity offset of broad emission lines relative to the host. Using the SDSS DR16 quasar catalogue, we estimate the velocity offset, $\Delta v$, as the difference between the broad H$\beta$ redshift and a noise-weighted redshift from narrow lines ([O\,III] 5007, [O\,II] 3728, and Ca\,II 3934). We adopt the redshift-relative colour excess $\Delta(g-i)$ as a proxy for dust column density. Analysing $\sim10^{5}$ quasars that meet basic spectral quality requirements, we find a modest but highly significant positive correlation between $|\Delta v|$ and $\Delta(g-i)$ (Spearman $r\simeq0.12$ and Pearson $r\simeq0.13$, with $p\ll10^{-10}$ in both cases). The fraction of highly obscured quasars increases with $|\Delta v|$, indicating that the correlation is driven by a dust-reddened subpopulation. The result is robust to the choice of minimum $|\Delta v|$ threshold and to the line redshift estimator (peak vs.~centroid). As expected, the correlation is largely absent when velocity offsets are computed between narrow emission lines. We find systematic differences between redshifted and blueshifted subsamples, which may point to residual velocity biases or additional physical effects (e.g.~winds, inflows, orientation-dependent obscuration, or asymmetric broad-line regions). Recoiling massive black holes provide a natural explanation for the observed correlation, but alternative scenarios should be explored. If confirmed, this would enable population-level constraints on massive black hole merger rates, recoil dynamics, and active galactic nuclei disc properties.
\end{abstract}

\begin{keywords}
black hole physics -- methods: observational -- methods: statistical -- galaxies: active -- galaxies: nuclei
\end{keywords}



\section{Introduction}
\label{sec:intro}

Observational evidence indicates that nearly all massive galaxies host a supermassive black hole (SMBH) at their centres~\citep{Richstone1998,KormendyHo2013}. These can form gravitationally bound SMBH binaries after galaxy mergers, which occur frequently in hierarchical structure formation~\citep{Begelman_Blandford_Rees1980,Lacey_Cole1993, Volonteri_Haardt_Madau2003}. This idea is also corroborated by observational evidence of spectroscopic (see e.g.~\citealt{Comeford_et_al2009,SMBHB_SDSS_Eracleous2012}) and photometric~\citep{CRTS_candidates_Graham_et_al2015, PTF_candidates_Charisi_et_al2016} binary candidates (see also recent reviews by~\citealt{Bogdanociv_et_al_LRR2022,Dorazio_Charisi_EMSigReview2023}). These binaries are also the most natural source of the recently observed stochastic gravitational-wave background at nanohertz frequencies (see e.g.~\citealt{nanograv_15yr_astro, epta_dr2_interp}).

Many SMBH binaries are expected to merge after inspiralling due to gravitational-wave emission~\citep{Begelman_Blandford_Rees1980, Milosavljevi_Merritt2001, Sesana_at_al2005}, and will be observable in the mHz gravitational-wave band by LISA \citep{LISAAmaro-Seoane2017, LISA_SMBHB_Klein2016}. The merger remnant SMBH can receive a recoil velocity on the order of $100-1000\ \mathrm{km/s}$ due to anisotropic gravitational-wave emission, depending on the binary mass ratio and spin orientation (see \citealt{black_hole_recoil_Healy2014} and references therein). This recoil can induce an oscillatory motion of the remnant SMBH, damped by dynamical friction~\citep{recoil_Madau_Quataert2004, oscill_Momossa_Merritt2008, Tanaka_Haiman2009}. It is expected that such a SMBH can retain the inner region of its accretion disc and remain active for $10^7-10^8$ years after the merger~\citep{EjectedQSO_timescale_Loeb2007}, allowing it to be observable as a quasar (QSO) with detectable kinematic and spatial signatures of the merger~\citep{RecoilingBH_Bonning_et_al2007, LippaiFreiHaiman2008, RecoilingBH_Blecha_et_al2011, RecoilingBH_Komossa2012, RecoilingBH_Blecha_et_al2016}. 

Although individual recoiling SMBH candidates can be identified (see e.g.~\citealt{RecoilingBH_Komossa2012}), such searches face challenges analogous to those in binary searches, in particular the difficulty of distinguishing genuine signals from alternative astrophysical explanations and avoiding false positives (see e.g.~\citealt{offset_detectability_Kelley2021}).
Some promising candidates include the spatially offset QSO in 3C~186, displaced by $\sim11$~kpc from the host-galaxy centre \citep{Chiaberge2017}; the quasar SDSS J092712.65+294344.0, which shows broad lines blueshifted by $\sim 2650 \ \mathrm{km/s}$ relative to narrow lines \citep{Komossa2008}; the system SDSS J1056+5516, which has been interpreted either as a triple active galactic nucleus (AGN) system or as a recoil candidate \citep{Kalfountzou2017}; the spectro-astrometrically displaced QSOs in E1821+643 \citep{Robinson2010,Jadhav2021}; and 2MASX J00423991+3017515, which combines a spatial offset of $\sim 3.8$ kpc with broad Balmer lines blueshifted by $\sim 1540\ \mathrm{km/s}$ \citep{Hogg2021}. Another candidate, CID-42 \citep{BlechaCID42_2013,NovakCID42_2015}, has recently been challenged by JWST imaging, which favours a merging galaxy pair in which only the south-eastern component hosts an unobscured AGN \citep{Li2024}. More recently, systematic integral-field surveys have begun to reveal off-nuclear broad-line AGNs consistent with moderate recoil velocities \citep{Barrows2025}.

A complementary, population-based statistical approach was proposed by \citet{AGNOsc_theory2016}, which probes the entire population of recoiling SMBHs rather than focusing on individual candidates. They demonstrated through simulations that a population of recoiling SMBHs should lead to a positive correlation between the obscuring dust column density, $\Sigma_{\rm dust}$, and the magnitude of the line-of-sight velocity of the SMBH relative to the obscuring torus, $|v_{\rm SMBH}|$. Physically, this correlation arises because gravitational recoil displaces SMBHs from the galactic centre, causing them to escape or execute damped oscillations with amplitudes comparable to or larger than the size of the dusty torus. Since the SMBH's velocity is largest when it is closest to the centre, the line of sight to higher-velocity SMBHs more frequently intersects larger dust columns. Although the tightness of the correlation and the subset of the QSO population showing the correlation can vary with different models, the presence of the correlation was found to be robust against model choice and measurement errors. Comparing with observations also allows testing the assumptions of these simulations, such as SMBH recoil, subsequent trajectory, and dust tori models.

In this paper, we use the SDSS-DR16 QSO catalogue~\citep{SDSS-DR16Q} to conduct the first search for a correlation between $\Sigma_{\rm dust}$ and $|v_{\rm SMBH}|$. Although the catalogue does not explicitly contain these quantities, it provides data products that serve as good proxies~\citep{AGNOsc_proc2017}. We use the $(g-i)$ colour excess of QSOs compared to other QSOs at the same redshift, $\Delta(g-i)$, as a proxy for $\Sigma_{\rm dust}$ (see e.g.~\citealt{Sigma_dust_proxy_Ledoux_et_al2015}). The broad line region (BLR), bound to the SMBH, track its movement post-recoil, while narrow lines from more distant narrow line regions (NLRs) should remain unaffected. Thus $v_{\rm SMBH}$ can be estimated as the velocity of the BLR relative to the NLR, $\Delta v$, from the offset between the broad and narrow lines in the QSO spectrum \citep{RecoilingBH_Bonning_et_al2007}.

This paper is organised as follows. In Section~\ref{sec:DA} we discuss the data and methodology we use, including quality cuts we imposed on the sample of QSOs analysed. In Section~\ref{sec:Results} we present our main results, showing a correlation between $\Delta(g-i)$ and $|\Delta v|$. In Section~\ref{sec:checks} we describe several checks we have done, and discuss what these mean for possible interpretations of our result. Finally, we provide concluding remarks and outline possible future directions in Section~\ref{sec:Conclusions}.

\section{Data and Methods}
\label{sec:DA}
Our analysis uses the the SDSS-DR16 QSO catalogue~\citep{SDSS-DR16Q} which contains photometric and spectroscopic measurements of 750,414 QSOs. We also make use of the catalogue of emission line properties based on SDSS-DR16 presented in \citet{Wu_Shen2022}. To look for the correlation proposed by \citet{AGNOsc_theory2016} we need proxies for $\Sigma_{\rm dust}$ and $v_{\rm SMBH}$, which can be derived from quantities available in these catalogues. In Section~\ref{ssec:Dv}, we describe how we achieve this for the line-of-sight velocity of the SMBH, $v_{\rm SMBH}$, and in Section~\ref{ssec:dgmr}, we discuss a suitable proxy for the dust column density, $\Sigma_{\rm dust}$. Finally, we describe how we cleaned our large sample of QSOs with various quality and consistency cuts in Section~\ref{ssec:quality_cuts}.

\subsection{Radial velocity}
\label{ssec:Dv}

In what follows, we assume that a recoiling SMBH will move together with its BLR\footnote{The conditions under which this is not expected to happen are discussed in Section~\ref{ssec:quality_cuts}.}, while the NLR will be left behind and exhibit a redshift consistent with that of the host galaxy and the dust torus at the galactic centre. Then the line-of-sight velocity of the SMBH relative to the obscuring dust torus ($v_{\rm SMBH}$) can be approximated with the line-of-sight velocity of the BLR w.r.t.~the NLR ($\Delta v$), which can be expressed as
\begin{equation}
v_{\rm SMBH} \approx \Delta v = \frac{z_{\mathrm B} - z_{\mathrm N}}{1 + z_{\mathrm N}} c,
\label{eq:dv}
\end{equation}
where $z_{\mathrm B}$ and $z_{\mathrm N}$ are the redshifts of any suitable broad and narrow lines, respectively, and $c$ is the speed of light. The selection of appropriate broad and narrow lines is of crucial importance. We need to be confident that the broad (narrow) line is indeed emitted primarily in the BLR (NLR) in order for $\Delta v$ to represent the relative velocity of the two.

In our analysis we use spectral fit results of 750,414 SDSS-DR16 QSOs available from \citet{Wu_Shen2022}. They provide raw fitted parameters for a large number of emission lines, and they also provide systemic redshift values ($z_{\rm sys}$) for seven specific lines: H$\beta$, Mg II, C III], C IV, [O III] 5007, Ca II 3934, and [OII] 3728\footnote{Note that while Si IV is listed in \citet{Wu_Shen2022}, its data is not available in the data release.}. These $z_{\rm sys}$ redshifts have been calibrated for known systematic offsets of these lines, some of which are correlated with QSO luminosity \citep{Shen2016_velocity_shifts}.

From the seven lines above, the one with the clearest association with the BLR is H$\beta$ \citep{Bentz2013HBeta}. \citet{Wu_Shen2022} fit this line with both broad and narrow components. The reported H$\beta_{\rm br}$ redshift is calculated from the broad components only. Thus we will equate $z_{\mathrm B}$ with the redshift of H$\beta_{\rm br}$ in our analysis. Extending our analysis to additional lines will be investigated in future work, and could provide invaluable insight into whether our results hold for other lines tracing the BLR.

We will use three of the above lines for calculating $z_{\mathrm N}$: [O III] 5007, Ca II 3934, and [OII] 3728. The two oxygen lines are forbidden emission lines and therefore originate in the low-density gas of the NLR, rather than the BLR \citep{KrolikAGNBook1999}. Ca II 3934 is a stellar absorption line from the host galaxy \citep{HewettWildQSOz2010}, so in principle, Ca II should be the most reliable indicator of the host galaxy redshift. However, there are two important advantages to including the oxygen lines. First, Ca II is significantly weaker than [O II] and [O III], leading to larger redshift uncertainties, and it is not measurable in roughly half of the QSOs in our sample. Second, although Ca II traces the host galaxy redshift, our goal is to estimate the systemic redshift of the central engine (i.e. the dust torus). The NLR, traced by the oxygen lines, may provide a comparably good (or potentially better) proxy for this quantity due to its closer physical association with the nucleus, despite possible kinematic offsets.

In practice, we determine $z_{\mathrm N}$ to be used in Eq.~(\ref{eq:dv}) as a noise-weighted average of all of the three narrow lines ([O III] 5007, Ca II 3934, and [OII] 3728) available for a given QSO:
\begin{equation}
    z_{\mathrm N} = 
    \frac{\sum_i z_i/\sigma_{z,i}^2}{\sum_i 1/\sigma_{z,i}^2},
\end{equation}
where $z_i$ and $\sigma_{z,i}$ are the redshift and corresponding uncertainty measured from the $i$th narrow line. This weighting ensures that more precisely measured lines dominate the final value of $z_{\mathrm N}$. This is the same procedure as the one used by \citet{Wu_Shen2022} to calculate their overall $z_{\rm sys}$, except that we only include the three narrow lines listed above instead of all seven lines. Note that some QSOs may not have all three of these available due to the quality cuts we impose (see Section~\ref{ssec:quality_cuts}) or simply due to the finite wavelength coverage of SDSS.  This method of determining $z_{\mathrm N}$ has the benefit over simply using the overall $z_{\rm sys}$ reported in \citet{Wu_Shen2022} that our redshift value is guaranteed to not be contaminated by any broad lines associated with the BLR.

\subsection{Dust column density}
\label{ssec:dgmr}

We follow the method put forward by \citet{AGNOsc_proc2017} to use the $g-i$ relative colour as a proxy for dust column density, which is defined as
\begin{equation}
    \Delta (g-i) \equiv (g-i)_{\mathrm{obs}} - \left\langle (g-i)_{\mathrm{obs}} \right\rangle_{z} , 
\end{equation}
where $(g-i)_{\mathrm{obs}}$ is the Galactic extinction corrected $g-i$ colour index, and $\left\langle  \right\rangle_{z}$ indicates averaging over QSOs in the same redshift bin. QSO colours evolve strongly with redshift due to the shifting of broad emission lines and continuum features through the photometric bands. As a result, a simple colour index such as $g-i$ cannot be interpreted directly as a reddening indicator across a wide redshift range. \citet{dgmiRichards2003} showed that subtracting the mean QSO colour as a function of redshift removes this systematic evolution, producing a relative colour $\Delta(g-i)$ that isolates object-to-object deviations from the typical QSO spectral energy distribution. In the SDSS QSO population the distribution of relative colours is approximately Gaussian on the blue side but exhibits a pronounced red tail, which is naturally explained by dust reddening. This approach therefore allows the identification of reddened QSOs in large heterogeneous samples without requiring detailed spectral modelling.

The usefulness of $\Delta(g-i)$ as a reddening proxy is further supported by the fact that the $g-i$ colour of QSOs is dominated by the QSO power-law continuum and line emission rather than host-galaxy light or Ly$\alpha$  forest absorption over a wide range of redshifts ($0.6\lesssim z \lesssim 2.2$). \citet{dgmiRichards2003} demonstrated that $\Delta(g-i)$ separates intrinsically red (steep-spectrum) QSOs from those whose colours cannot be explained by a simple power-law continuum and instead require dust reddening. In practice, QSOs with sufficiently large positive colour excess (typically $\Delta(g-i)\gtrsim0.3-0.5$ depending on redshift) are interpreted to be significantly dust reddened rather than simply exhibiting intrinsic continuum slope variations.

$\Delta (g-i)$ values were readily available in the SDSS DR12 QSO catalogue \citep{SDSS-DR12Q}, but are not present in DR16 \citep{SDSS-DR16Q}. As a result, we calculate our own $\Delta (g-i)$ values, which we validate on a DR12 sample to produce consistent results. We set up our redshift bins such that each bin has one thousand QSOs. This results in $\Delta (g-i)$ standard errors of less than 0.02 for all QSOs, and less than 0.01 for most. We emphasize that these uncertainties reflect only the statistical error on the mean reddening estimated within each redshift bin, and do not include the photometric measurement errors on individual $g$ and $i$ band magnitudes, nor the effect of the mean changing within a redshift bin. The former is typically expected to be smaller than this error, and the latter is partially mitigated by the fact that we use a linear interpolation between the mean redshifts of two adjacent bins to calculate the mean reddening at any given redshift. Note that while we use only a subset of QSOs in SDSS-DR16 QSO catalogue for our analysis (see Section~\ref{ssec:quality_cuts}), we calculate $\Delta (g-i)$ values on the full set of QSOs since these values depend on the entire population of QSOs at a given redshift. We excluded only 1,001 QSOs from this calculation due to the lack of available $g$ or $i$ band magnitudes.

\subsection{Quality cuts}
\label{ssec:quality_cuts}

To obtain a clean sample of high-quality QSO spectra relevant to recoiling SMBHs, we applied several quality cuts to the full sample of 750,414 QSOs in SDSS-DR16 QSO catalogue. The numbers of QSOs remaining after various cuts are summarised in Table~\ref{tab:qso_numbers}. Note that these numbers depend on what line we consider as the broad line. As we discussed in Section~\ref{ssec:Dv}, we use H$\beta_{\rm br}$ as the broad line in our main analysis, but we also perform null tests where one of our three narrow lines is treated as the broad line (see Section~\ref{ssec:other_lines}), so Table~\ref{tab:qso_numbers} lists QSO numbers with these choices as well.

The majority of the QSOs are filtered out by the trivial requirement that the broad line and at least one narrow line needs to be in band at the given QSOs redshift. We also filter out all QSOs with a $z_{\rm sys}$ value of -1 or -2, as these indicate unreliable fits. QSOs where either $g$ or $i$ band magnitude is not available are also filtered out as these are needed to calculate $\Delta (g-i)$. Beyond these strictly necessary filters, we also apply quality cuts proposed by \citet{Wu_Shen2022} that require a minimum line flux signal-to-noise ratio of 2; a line flux between 10$^{38}$ erg/s and 10$^{48}$ erg/s; that at least half of the pixels are available in the given line complex; and that the reduced $\chi^2$ of the fit is between 0 and 2. We also filter out a small number of QSOs where our $z_{\mathrm N}$ derived from the three narrow lines detailed above is inconsistent at the $3\sigma$-level with the systemic redshift derived by \citet{Wu_Shen2022} from both broad and narrow lines. We collectively call all these filters "basic quality cuts" in Table~\ref{tab:qso_numbers}.

We also make further cuts based on the resulting $\Delta v$ peculiar velocities. We filter out all QSOs with $|\Delta v|>2700$ km/s, because allowing higher velocities results in an artificial excess of values due to the line reaching the edge of the fitted wavelength window. This value is chosen to conservatively exclude sources close to the H$\beta_{\rm br}$ centroid fitting boundary of approximately 3000 km/s used by \citet{Wu_Shen2022}. In addition, we also filter out QSOs with particularly small peculiar velocities, $|\Delta v|<45$ km/s. These are typically consistent with $\Delta v=0$ within errors, and would be dominated by single non-recoiling SMBHs, which would dilute the correlation. The threshold was chosen as the one that would result in the strongest correlation based on theoretical results from \citet{AGNOsc_theory2016}. However, the results are insensitive to this threshold, as we show in Section~\ref{sec:Results}. We collectively call these "$|\Delta v|$ outlier cuts" in Table~\ref{tab:qso_numbers}.

Finally, we apply a physically motivated filter that ensures that the measured $|\Delta v|$ and Gaussian line width of the broad line ($\sigma_B$) are consistent with the interpretation where the BLR remains bound to the SMBH and follows it as it recoils \citep{LippaiFreiHaiman2008}. While we only have access to the line-of-sight velocity, assuming an isotropic distribution of recoil directions, we can impose the order of magnitude constraint $\sigma_B>|\Delta v|$. We call this the "broad line consistency cut" in Table~\ref{tab:qso_numbers}. If this condition is not satisfied, we do not expect the BLR to remain bound to the SMBH, therefore, our measured $\Delta v$ cannot correspond to the velocity of the SMBH w.r.t.~the central engine, $v_{\rm SMBH}$. We can see that this filter removes a small number of QSOs compared to the basic quality cuts. We will show in Section~\ref{sec:checks} that our results are qualitatively unchanged if we do not apply this filter.

\begin{table}
	\centering
	\caption{Number of QSOs with different lines available after various quality and consistency cuts (total number in the SDSS DR16 QSO catalogue: 750,414). The largest reduction in QSO numbers is due to the line in question being out of band. See details in Sec.~\ref{ssec:quality_cuts}.}
	\label{tab:qso_numbers}
	\begin{tabular}{lcccc} 
		\hline
		 & H$\beta_{\rm br}$ & [O III] & [O II] & Ca II\\
         Rest-frame wavelength [\AA] & 4861 & 5007 & 3728 & 3934\\
		\hline
		After basic quality cuts         & 123,160 & 105,068 & 117,577 & 73,045\\
        After $|\Delta v|$ outlier cuts   & 109,711 & 70,265 & 79,654 & 60,793\\
		After BLR consistency cut$^a$ & \textbf{104,688} & \textbf{58,090} & \textbf{62,978} & \textbf{41,937}\\
		\hline
        \multicolumn{5}{l}{\footnotesize$^a$ This represents the final number of QSOs used for the analysis.}\\
	\end{tabular}
\end{table}

\section{Results}
\label{sec:Results}

\begin{figure}
\centering
\includegraphics[width=1\linewidth]{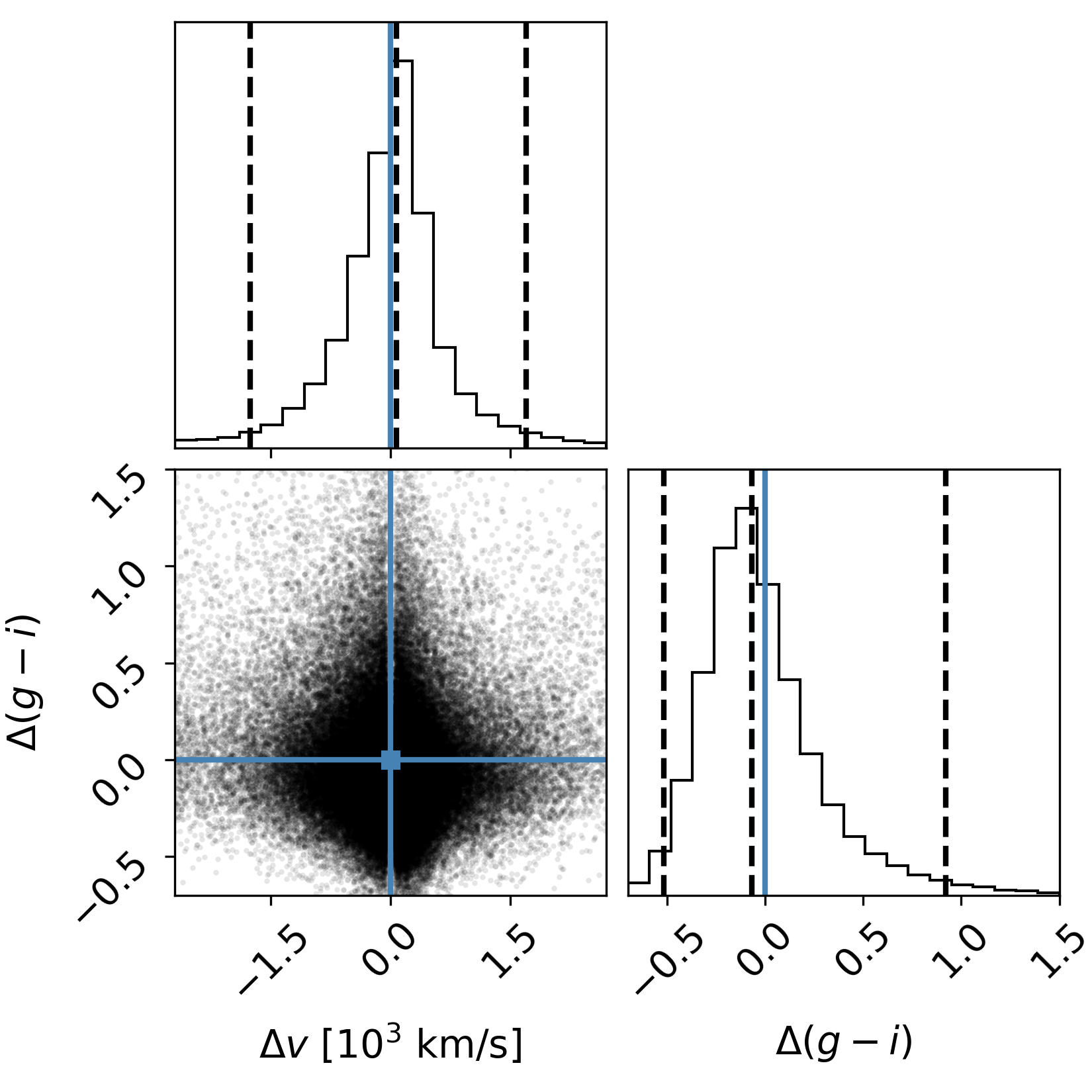}
\caption{Distribution of $\Delta v$ and $\Delta (g-i)$ values for $\sim$100,000 QSOs with available H$\beta_{\rm br}$ after quality and consistency cuts (see Table~\ref{tab:qso_numbers}). The blue point and lines mark zero values of both parameters. Dashed lines show the 2nd, 50th, and 98th percentiles (corresponding to a 2-$\sigma$ range). Note that while the distribution of $\Delta v$ is quite symmetric, the distribution of $\Delta (g-i)$ shows a heavy tail of positive values, corresponding to a population of dust-reddened QSOs.}
\label{fig:Dv_DGMR_corner}
\end{figure}

Figure~\ref{fig:Dv_DGMR_corner} shows the distribution of $\Delta v$ (from H$\beta_{\rm br}$) and $\Delta (g-i)$ for the $\sim$100,000 QSOs remaining after all the various cuts described in Section~\ref{ssec:quality_cuts}. The blue marker indicates $\Delta v=0$, $\Delta (g-i)=0$. Dashed lines indicate the median, 2nd and 98th percentiles, roughly corresponding to a 2-$\sigma$ region. We can see that the distribution of $\Delta v$ is symmetric and centred roughly on zero, with a slight preference towards positive values (median $\Delta v$ of $\sim$70 km/s) corresponding to systems where the BLR is redshifted compared to the NLR. The distribution of $\Delta (g-i)$ is also centred on zero, but exhibits a heavy tail towards positive values. This is similar to the results found, e.g., in~\citet{dgmiRichards2003} and can be attributed to the presence of dust-reddened QSOs exhibiting large values of $\Delta (g-i)$.

\begin{figure*}
\centering
\includegraphics[width=0.85\textwidth]{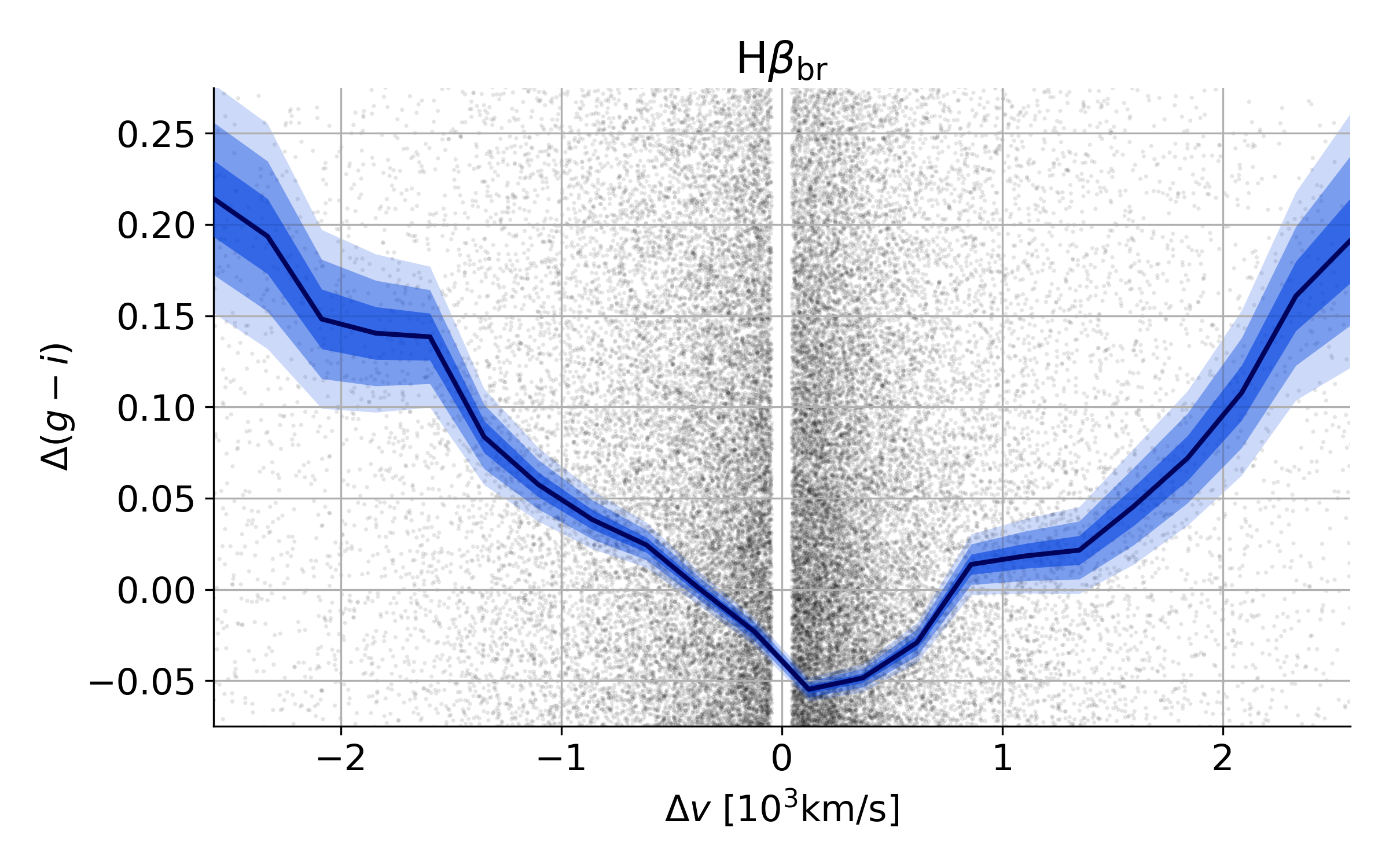}
\caption{$\Delta (g-i)$ relative colour (proxy for dust reddening) as a function of the line-of-sight peculiar velocity of the BLR relative to the NLR, $\Delta v$, obtained from the broad H$\beta$ emission line. Black dots correspond to $\sim$100,000 individual QSOs after various quality cuts (see Tab.~\ref{tab:qso_numbers} and Sec.~\ref{ssec:quality_cuts}). The blue line and bands show average values in 22 equally spaced $\Delta v$ bins and their corresponding 1/2/3-$\sigma$ errors. These show a clear positive correlation between $\Delta (g-i)$ and the magnitude of the line-of-sight velocity $|\Delta v|$. The band with no QSOs around $\Delta v = 0$ is because we filter out all QSOs with $|\Delta v|<45$ km/s (see Sec.~\ref{ssec:quality_cuts}).}
\label{fig:dv_vs_dgmr}
\end{figure*}

\citet{AGNOsc_theory2016} predicts a correlation between dust reddening and the magnitude of the BLR radial velocity. To investigate this correlation, we show $\Delta v$ and $\Delta (g-i)$ (proxy for dust reddening) in Figure~\ref{fig:dv_vs_dgmr} as grey dots. We also show the mean $\Delta (g-i)$ values, along with 1$\sigma$, 2$\sigma$ and 3$\sigma$ errors on the mean, in 22 equal-width radial velocity bins. While the individual data points are not visibly correlated, the binned data shows a clear trend, where QSOs with both large positive (redshifted) and large negative (blueshifted) velocities tend to have higher dust reddening than those with $\Delta v\simeq0$. This is as expected if the sample contains BLRs associated with recoiling SMBHs \citep{AGNOsc_theory2016}. To quantify the putative correlation between $|\Delta v|$ and $\Delta (g-i)$, we calculate the Spearman correlation coefficient between these quantities. This correlates the rank of values in an ordered list, so unlike the Pearson correlation, it does not assume a linear relationship. Applying this to the unbinned data we find a correlation coefficient of $r=0.118$, and a $p$-value of $p\ll10^{-10}$, indicating a modest, but highly significant correlation\footnote{Note that whenever the $p$-value is smaller than $10^{-10}$ we do not quote its exact value, as these all indicate high significance, and the exact value is sensitive to small changes in the included QSOs and other modelling choices.}. We also find a comparable result with a Pearson correlation test, yielding $r=0.133$, and a $p$-value of $p\ll10^{-10}$. This is not surprising given the approximately linear relationship seen in the mean values shown in Fig.~\ref{fig:dv_vs_dgmr}.

Most importantly, the sign of the correlation agrees with the expectation, and the value of the Pearson correlation coefficient ($r\simeq0.133$) is in excellent agreement with that predicted in \citet{AGNOsc_theory2016} for partially obscured QSOs with perfectly measured velocities ($r\simeq0.13$) and is slightly lower than the value predicted for the entire population ($r\simeq0.28$). The slightly lower correlation coefficient can be attributed to measurement errors that dilute the correlation. In fact, \citet{AGNOsc_theory2016} found that statistical errors of $\pm100$ km/s can reduce the correlation coefficient to $r\simeq0.03$. In addition, the correlation can also be diluted if only a fraction of the QSO population is associated with a recent merger event. In fact, this can in principle be used in the future to estimate the fraction of QSOs with recent SMBH mergers, as we outline in Section~\ref{sec:Conclusions}.

Note that \citet{AGNOsc_theory2016} tested the correlation between $\log_{10}(|v_{\rm SMBH}|)$ and $\Sigma_{\rm dust}$, not between $|\Delta v|$ and $\Delta (g-i)$ as we do here. However, the above comparison is still valid given that $\Delta v$ is meant to approximate $v_{\rm SMBH}$, and $\Delta (g-i)$ is a linear function of $\Sigma_{\rm dust}$. The only concern could be the presence of the logarithm, however we find very similar correlation results between $\log_{10}(|\Delta v|)$ and $\Delta (g-i)$ with a Pearson $r=0.112$\footnote{Note that the Spearman $r$ value is unchanged by the logarithm as this test is insensitive to monotonic transformations by design.}.

Although Fig.~\ref{fig:dv_vs_dgmr} clearly shows that the mean $\Delta (g-i)$ increases with $|\Delta v|$, it is not clear whether this is driven by dust-reddened QSOs. Even in the absence of dust reddening, QSOs can exhibit a wide range of $\Delta (g-i)$ values due to intrinsic variations in the slope of their power-law continuum spectra. However, objects with sufficiently large $\Delta (g-i)$ values are unlikely to be explained by intrinsic spectral differences alone and can be more confidently classified as dust reddened. \citet{dgmiRichards2003} defines dust-reddened QSOs with thresholds of $\sim0.3-0.5$ depending on redshift. For simplicity, we adopt redshift-independent criterion of the form $\Delta (g-i) > \kappa$, where $\kappa \simeq 0.0$–$0.5$. Increasing $\kappa$ reduces contamination from intrinsically red QSOs, at the cost of excluding a larger fraction of genuinely dust-reddened objects. 

\begin{figure}
\centering
\includegraphics[width=\linewidth]{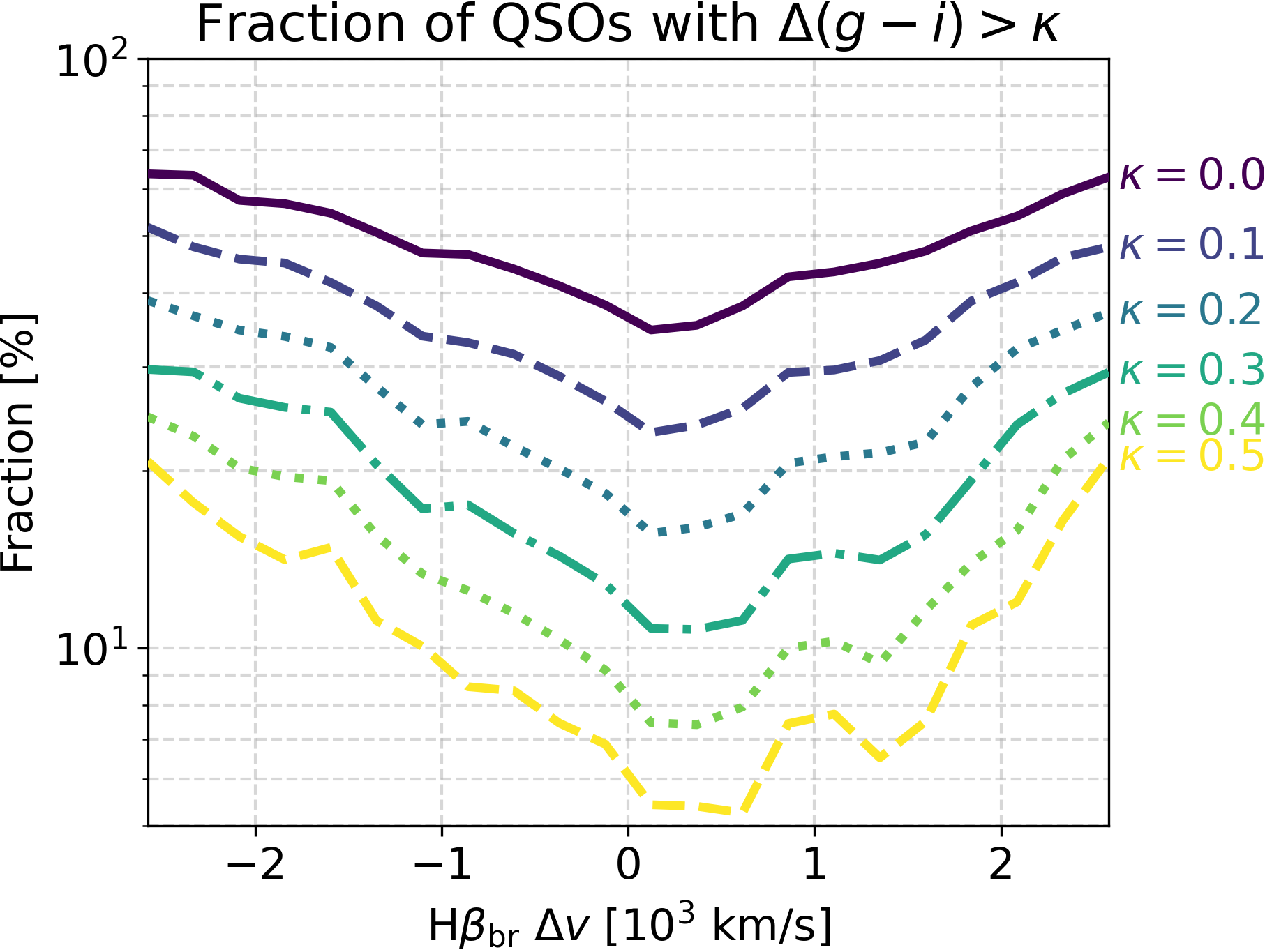}
\caption{Fraction of dust-reddened QSOs with $\Delta (g-i)$ above various $\kappa$ thresholds as a function of H$\beta$-derived $\Delta v$. These show a similar trend as the mean $\Delta (g-i)$ (cf.~Fig.~\ref{fig:dv_vs_dgmr}), with increasing fraction of dust-reddened QSOs at large $|\Delta v|$ values. Notice that the trend is stronger for higher thresholds, which suggests that the effect is indeed due to dust-reddened QSOs as opposed to ones with an intrinsically steep power-law continuum.}
\label{fig:red_qso_fraction}
\end{figure}

Figure~\ref{fig:red_qso_fraction} shows the fraction of dust-reddened QSOs using various thresholds as a function of $\Delta v$. We can see that QSOs with higher line-of-sight BLR velocities are more often dust reddened. The effect is particularly pronounced when using higher thresholds, with the fraction increasing from about 6\% (35\%) at zero velocity to about 20\% (65\%) at 2700 km/s for a threshold value of $\kappa = 0.5$ ($\kappa = 0.0$), a roughly three-fold (two-fold) increase. This behaviour can be reconciled with Fig.~\ref{fig:dv_vs_dgmr}, where the mean $\Delta (g-i)$ remains in the range $\sim -0.05$–$0.2$, i.e.~below typical dust-reddening thresholds unless very low values of $\kappa$ are adopted. Together, these results indicate that dust-reddened QSOs constitute only a minority of the overall population, especially at higher $\kappa$, but that this subpopulation becomes more prevalent at larger $|\Delta v|$. As a result, a relatively small fraction of objects with high $\Delta (g-i)$ values is sufficient to drive the increase in the mean, implying that the observed correlation is indeed primarily due to dust reddening rather than changes in the intrinsic continuum slope.

\begin{figure}
\centering
\includegraphics[width=\linewidth]{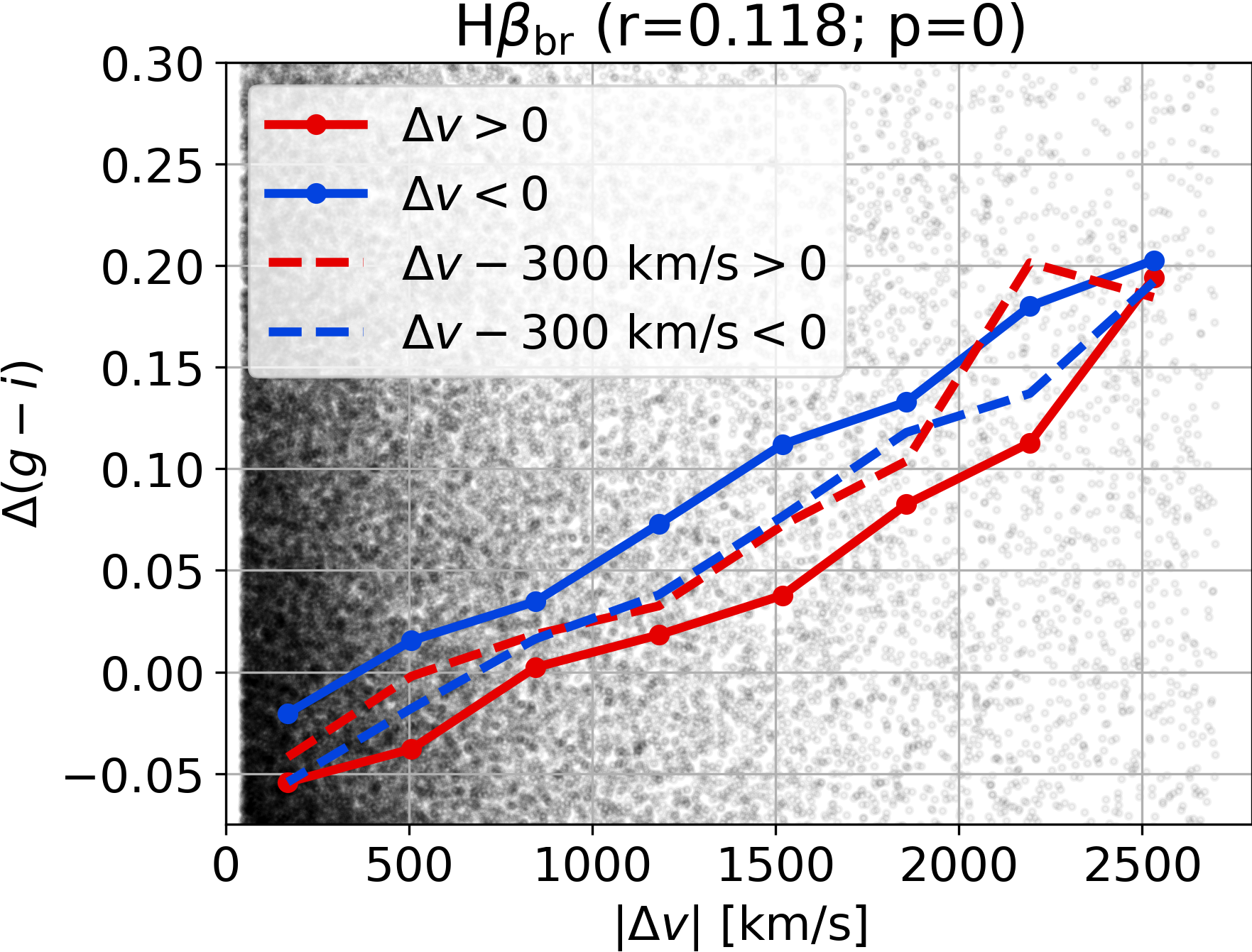}
\caption{$\Delta (g-i)$ relative colour (proxy for dust reddening) as a function of the absolute value of the line-of-sight peculiar velocity of the BLR relative to the NLR, $|\Delta v|$, obtained from the broad H$\beta$ emission line (i.e.~a version of Fig.~\ref{fig:dv_vs_dgmr} folded in half along the horizontal axis). Black dots correspond to $\sim$100,000 individual QSOs after various quality cuts (see Table~\ref{tab:qso_numbers}). The solid blue (red) line shows binned average values for negative (positive) $\Delta v$ values. These show a clear positive correlation between $\Delta (g-i)$ and $|\Delta v|$, which is consistent with the results of a Spearman correlation test on the unbinned data yielding $r=0.118$ and $p\ll10^{-10}$. Note that redshifted and blueshifted BLRs show a systematic offset in $\Delta (g-i)$. The two populations can be brought into agreement by applying a uniform shift $\Delta v \rightarrow \Delta v - 300\ \mathrm{km/s}$, indicating that the measured velocities may be systematically biased toward redshifted values by $\sim$300 km/s.}
\label{fig:dv_vs_dgmr_shift}
\end{figure}

Although the positive correlation we find between $|\Delta v|$ and $\Delta (g-i)$ is consistent with the SMBH recoil scenario, what is unexpected is the slight asymmetry between the redshifted and blueshifted subsamples. This is readily visible in Fig.~\ref{fig:dv_vs_dgmr}, where the left hand side of the blue line is systematically higher than the right hand side. We can also see that the minimum of the curve is not at $\Delta v =0$, but is instead shifted towards positive values\footnote{A similar shift of the minimum towards positive $\Delta v$ values is also visible in Fig.~\ref{fig:red_qso_fraction}.}. To demonstrate this, we show $|\Delta v|$ vs.~$\Delta (g-i)$ separately for redshifted and blueshifted BLRs in Figure~\ref{fig:dv_vs_dgmr_shift}. We can see that the lines corresponding to positive (red) and negative (blue) line-of-sight velocities are offset from each other. This means that blueshifted BLRs tend to be more dust obscured than redshifted ones at a given line-of-sight velocity. This cannot be easily reconciled with the SMBH recoil scenario, where we would expect to see most recoiling SMBHs during their first oscillation as they leave the galactic nucleus. This means that redshifted BLRs should be behind the dust torus, thus exhibiting similar or stronger dust reddening than blueshifted BLRs.

One possible explanation could be if the radial velocity estimates have a systematic bias. To illustrate this, we show in Fig.~\ref{fig:dv_vs_dgmr_shift} how the binned data changes if we assume all $\Delta v$ values are overestimated by 300 km/s (dashed lines). This same shift also results in a slight increase in both the Spearman ($r=0.118\rightarrow 0.134$) and Pearson (${r=0.133\rightarrow0.146}$) correlation coefficients. While this is an arbitrarily chosen value, we can see that such a systematic bias could account for the asymmetry we see between redshifted and blueshifted BLRs. Although we do not know if such a bias is present in our sample, \citet{Shen2016_velocity_shifts} shows that H$\beta_{\rm br}$ line peaks tend to be shifted by $\sim$100 km/s w.r.t.~the host galaxy with an intrinsic scatter of $\sim$400 km/s. Although the redshift values we use from \citet{Wu_Shen2022} are already corrected for the shift, and the scatter in itself should not result in a systematic bias, this highlights that there are poorly understood systematics at the few 100 km/s level, which could be responsible for the asymmetry. 

Another possible explanation can come from asymmetric broad emission line profiles. These can conceivably manifest under the SMBH recoil scenario, since different parts of the BLR will be affected differently by the movement of the SMBH. If the line is asymmetric, it is no longer clear if either the mean or peak velocity of the profile will correspond to the velocity of the SMBH, so taking either of these as the SMBH velocity could lead to bias. We also expect any resulting line asymmetry to depend on the direction of the recoil, so the sign of the bias will depend on the sign of the line-of-sight velocity, $\Delta v$. This suggests that such a line asymmetry might explain the differences between redshifted and blueshifted BLRs we see in Figure~\ref{fig:dv_vs_dgmr_shift}. This is investigated in more detail in Section~\ref{ssec:v_sys_peak_cent}.

We also find a significant correlation between $|\Delta v|$ and the width of the H$\beta_{\rm br}$ line, characterised by the full width at half maximum (FWHM), with Spearman $r=0.327$, Pearson $r=0.334$, and $p\ll10^{-10}$ for both. This behaviour is expected under the recoiling SMBH interpretation. In order for the SMBH to retain its BLR, the recoil velocity must not exceed the velocity dispersion, i.e.~$|\Delta v|\lesssim \sigma_{\rm B}$ (see Section~\ref{ssec:quality_cuts}). This condition naturally introduces an upper bound on $|\Delta v|$ at a given FWHM, leading to an intrinsic correlation between these quantities. Importantly, this correlation is already present even without explicitly applying the $\sigma_{\rm B} = \rm{FWHM}/(2\sqrt{2\ln2}) > |\Delta v|$ consistency filter, indicating that our sample largely satisfies this physical constraint intrinsically. In fact, we see that even without the filter applied, our sample has very few QSOs violating this constraint (see Table \ref{tab:qso_numbers}), and they show an upper bound on $|\Delta v|$, which grows linearly with FWHM.

\begin{figure}
\centering
\includegraphics[width=\linewidth]{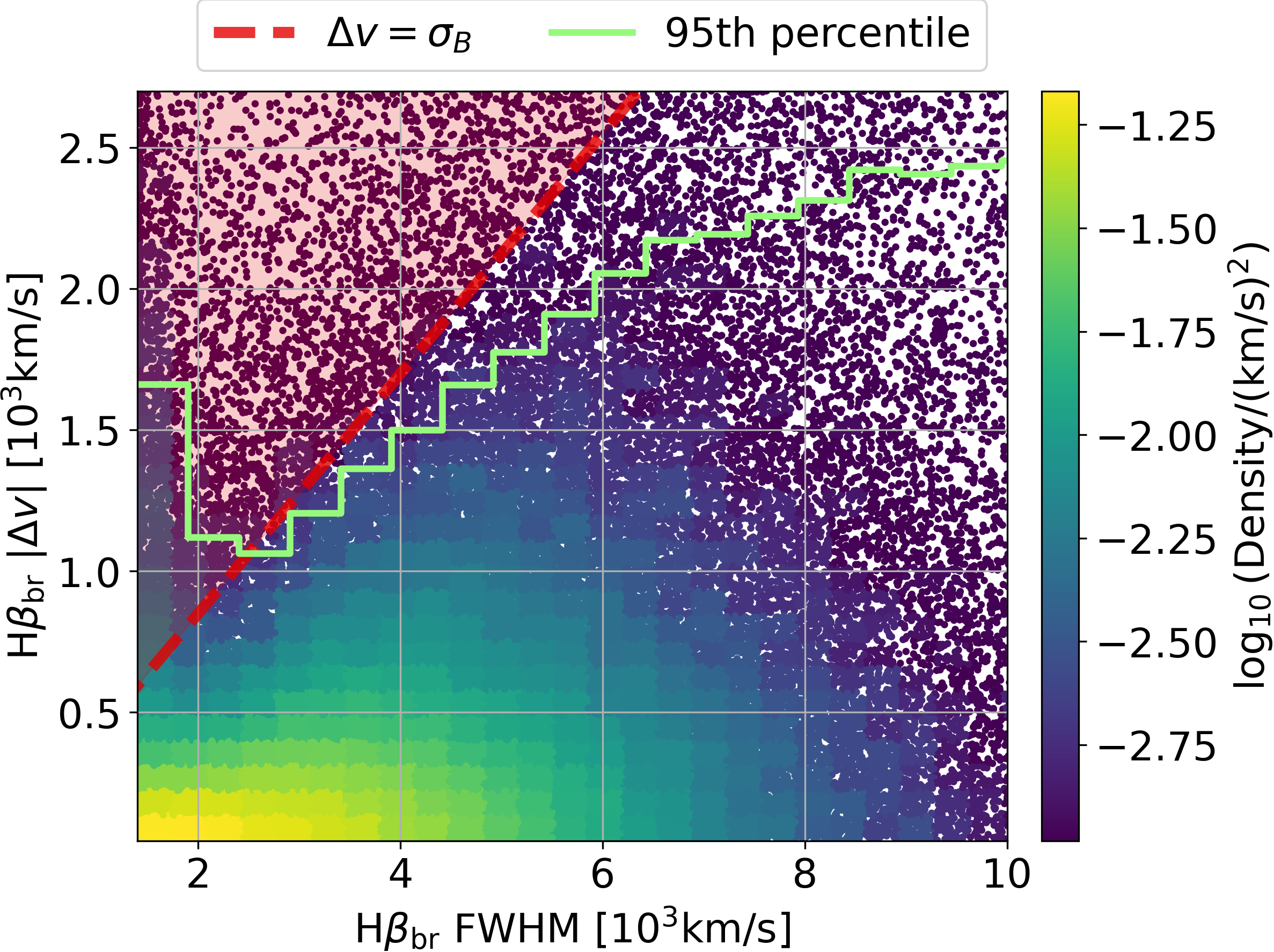}
\caption{H$\beta_{\rm br}$-derived $|\Delta v|$ values as a function of the FWHM of the H$\beta_{\rm br}$ line coloured with the density of QSOs on this $|\Delta v|-\rm{FWHM}$ plane. We also show the 95th percentile $|\Delta v|$ values in 17 equal-width FWHM bins (solid light green line). The dashed red line and shaded region above it marks the region where $|\Delta v|$ is too high compared to the FWHM for the BLR to remain bound to the SMBH. We can see that the distribution of QSOs largely respects this limit, except for a low number of outliers and an overdensity of QSOs at the lower FWHM bound, which is likely a fitting artifact (see text). This is consistent with the interpretation that a significant fraction of our QSO sample corresponds to recoiling SMBHs.}
\label{fig:FWHM_vs_Dv}
\end{figure}

Figure \ref{fig:FWHM_vs_Dv} illustrates this by showing the density of QSOs in the FWHM--$|\Delta v|$ plane. We can see that the distribution of $|\Delta v|$ values cuts off around the $|\Delta v|=\sigma_{\rm B}$ limit (red dashed line), with few QSOs above it. This is also visible from the 95th percentile of $|\Delta v|$ values at different FWMH values (light green), which also closely follows the $|\Delta v|=\sigma_{\rm B}$ line. The one exception is the overdensity of QSOs above the red line in the lowest FWHM bin. This feature is likely due to fitting artefacts, where the imposed lower bound on the FWHM of the broad-line component ($\sim1,400$ km/s) is reached. Below this limit, \citet{Wu_Shen2022} classify lines as narrow. These cases therefore likely correspond to spectra in which the broad and narrow components are not cleanly separable, leading the fitting routine to misattribute a narrow component to the broad-line model. The fact that our QSO sample naturally satisfies the limit required for the BLR to remain bound to the SMBH further supports our hypothesis that a significant fraction of our sample indeed represents recoiling QSOs. We also find a significant correlation between $|\Delta v|$ and the logarithm of the single-epoch SMBH mass estimated from the H$\beta_{\rm br}$ line \citep{virial_mass_Vestergaard2006,Shen2011_DR7,Shen2019}, with Spearman $r=0.242$, Pearson $r=0.196$, and $p\ll10^{-10}$ for both. This is likewise expected, as these mass estimates depend strongly on the FWHM of the broad emission line used \citep{Wu_Shen2022}, and therefore inherit the same underlying dependence on $|\Delta v|$.

\begin{figure}
\centering
\includegraphics[width=\linewidth]{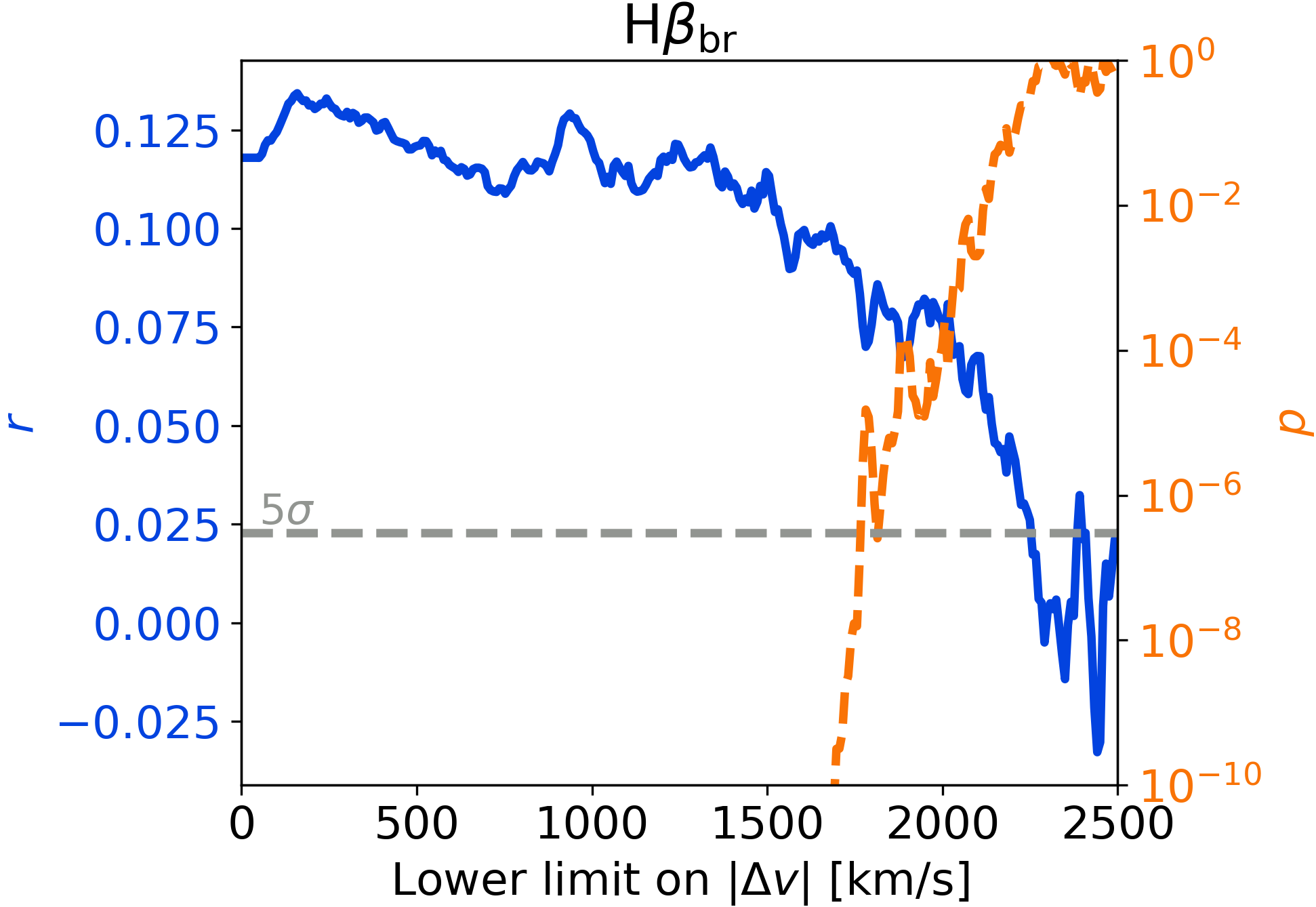}
\caption{Results of the Spearman correlation test as a function of the lower limit used on $|\Delta v|$. Blue solid line shows the $r$ correlation coefficient, and orange dashed line shows the $p$-value ($p<10^{-11}$ for $|\Delta v|\lesssim$1,600 km/s, where the orange line is not visible). Also shown is the $p$-value corresponding to $5\sigma$ significance (grey dashed line). Our main analysis used a limit of 45 km/s, but we can see that the correlation remains significant with $r=0.05-0.15$ for a large range of limits until about 2000 km/s. Above this, the majority of the QSOs get filtered out and the correlation is no longer significant.}
\label{fig:delta_v_limits}
\end{figure}

As described in Section \ref{ssec:quality_cuts}, our main correlation analysis was done on QSOs with $|\Delta v|>45$ km/s. This limit was chosen as the simulation results found the strongest correlation with this limit \citep{AGNOsc_theory2016}. Here we investigate how the correlation is affected by imposing different lower limits on $|\Delta v|$. Figure~\ref{fig:delta_v_limits} shows Spearman $r$ coefficients and $p$-values for different values of the $|\Delta v|$ lower cut. We can see that the presence of the correlation is robust against the choice of $|\Delta v|$ lower limit over a large range of values: it remains significant until the lower limit is increased to about 2000 km/s, at which point the correlation diminishes. This is not surprising given that a lower limit of 2000 km/s filters out more than 97\% of the QSOs. We also see that the correlation coefficient stays consistently $r=0.05-0.15$ over the same range of lower cut values. These results highlight that the correlation is not dominated by a few outliers or a particular subpopulation of QSOs in a specific $|\Delta v|$ range. The strongest correlation is found with a lower limit of $\sim 160$ km/s, which results in a Spearman correlation coefficient of $r\simeq0.134$. This is roughly consistent with the theoretical results, which found the strongest correlation with a limit of 45 km/s \citep{AGNOsc_theory2016}.

\section{Consistency Checks and Interpretation}
\label{sec:checks}

As we have seen in Section~\ref{sec:Results}, we find a significant correlation between the dust reddening of a QSO and the magnitude of the BLR radial velocity w.r.t.~the NLR. This can be explained by a population of recoiling SMBHs holding onto the BLR around them as proposed by \citet{AGNOsc_theory2016}. While this was our original motivation to search for such a correlation, we recognise that there could be other explanations for such an observational result. We discuss some of these alternative possibilities below.

Several physical processes can introduce large systematic velocity offsets between broad and narrow emission lines in AGN beyond a genuine bulk motion of the BLR \citep{Shen2016_velocity_shifts}. For example, radiatively driven winds in the BLR preferentially accelerate high‐ionisation gas toward the observer, producing blueshifts of lines such as C IV relative to low‐ionisation or narrow‐line tracers \citep{Gaskell1982,wind_emission_lines_Murray1995, CIV_wind_Richards2011}. It has also been suggested that light scattering off inflowing material can also result in blueshifting of high-ionisation lines \citep{inflow_Gaskell2016}. Low-ionisation lines like H$\beta$ are less affected by winds, motivating its use in our analysis. However, they can also have systematic shifts due to various reasons, e.g.~accretion disc rotation \citep{disk_Chen1989}, off-axis variability \citep{offaxis_var_Gaskell2011}, or orbital motion in a SMBH binary \citep{SMBHB_SDSS_Eracleous2012,SMBHB_SDSS_Ju2013,offset_BLR_SMBHBs_Liu2014}.

Narrow emission lines can also be shifted due to outflows (or inflows) driven by AGN feedback or host‐galaxy interactions, such that the “systemic” rest frame implied by the NLR may differ from the host stellar frame \citep{NRL_offset_Crenshaw2010}, and these offsets can depend on orientation \citep{NLR_inclination_Fischer2013}. Because any one of these mechanisms (or combinations thereof) can produce velocity offsets of a few hundred to a few thousand km/s, caution is required when interpreting observed BLR shifts as evidence of black‐hole motion. To help discern such alternative explanations, we performed additional tests and consistency checks, which we summarise below.

\subsection{Different ways to measure line-of-sight velocity}
\label{ssec:v_sys_peak_cent}

As described in Sec.~\ref{ssec:Dv}, our main analysis estimates the line-of-sight velocity of the BLR w.r.t.~the NLR by comparing the systemic redshift of H$\beta_{\rm br}$ with that of narrow lines. These systemic redshifts given by \citet{Wu_Shen2022} are calculated from the peak of the fitted line profile, and are corrected for population-average offsets of the given line from \citet{Shen2016_velocity_shifts}. Here we test how our results are affected by two changes: i) not applying corrections for line offsets; and ii) calculating redshifts from the centroids of line profiles instead of peak wavelengths. Although applying population based corrections is justifiable, the physical origin of these line shifts are not understood, and as we have shown in Fig.~\ref{fig:dv_vs_dgmr_shift}, a constant shift could explain the difference we see between redshifted and blueshifted BLRs. Also, as discussed in Section~\ref{sec:Results}, it is possible that a recoiling BLR would result in asymmetric line profiles, which means that the peak and centroid wavelengths could be significantly different, thus motivating the test of using one or the other.

\begin{figure}
\centering
\includegraphics[width=\linewidth]{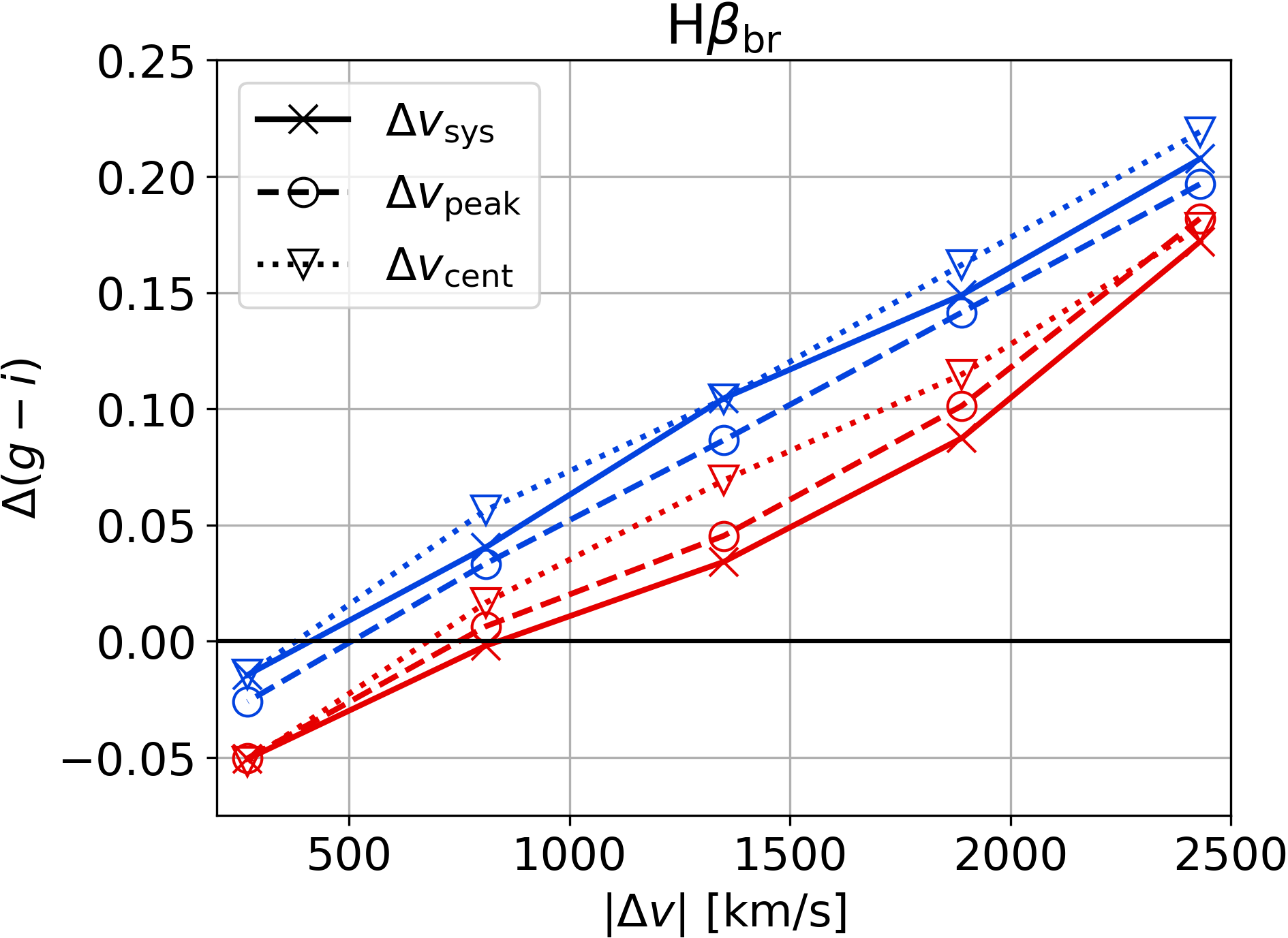}
\caption{$\Delta (g-i)$ relative colour as a function of the absolute value of the line-of-sight velocity of the BLR relative to the NLR, $|\Delta v|$, obtained from the broad H$\beta$ emission line. The solid blue (red) line shows binned average values for negative (positive) $\Delta v$ values obtained using the systemic line redshifts provided in \citet{Wu_Shen2022} ($\Delta v_{\rm sys}$). The dashed lines are the same, but based on raw peak wavelength values, which are not corrected for systematic shifts ($\Delta v_{\rm peak}$); and the dotted lines show results based on raw centroid wavelength values ($\Delta v_{\rm cent}$). Note that while the presence of the correlation is robust to these modelling choices, the amount of asymmetry between blueshifted and redshifted subsamples significantly depends on them.}
\label{fig:dv_sys_vs_peak_vs_cent}
\end{figure}

Figure~\ref{fig:dv_sys_vs_peak_vs_cent} shows the binned $|\Delta v|$ vs.~$\Delta (g-i)$ results using three different methods to calculate $\Delta v$: i) the systemic line redshifts as in our main analysis ($\Delta v_{\rm sys}$); ii) the raw peak wavelengths without correcting for systematic shifts ($\Delta v_{\rm peak}$); and iii) the raw centroid wavelengths without correcting for systematic shifts ($\Delta v_{\rm cent}$)\footnote{For simplicity, we have not tested the fourth possible combination of a shift-corrected redshift based on centroid wavelengths, as this would have required redoing the correction procedure of \citet{Wu_Shen2022}.}. We can see that while the results change somewhat, the qualitative result of a strong positive correlation remains. This highlights that the result is robust against the details of how the velocity offsets are estimated. Importantly, we see that the difference between the blueshifted and redshifted subsample is somewhat reduced in some of these alternative methods of calculating $\Delta v$. Note in particular that the two lines for $\Delta v_{\rm peak}$ move closer together compared to the lines for $\Delta v_{\rm sys}$. This indicates that some of the difference may be explained by the shift corrections and choice of best wavelength estimator, but even these alternative versions display differences between the redshifted and blueshifted samples.

Note, however, that the BLR, especially if it remains only marginally bound to the SMBH after its recoil, will likely be significantly distorted in its shape and in its kinematics.  These would introduce corresponding asymmetries and distortions in the broad emission line profiles from recoiling SMBHs. It is likely that neither the peak nor the centroid of these lines corresponds directly to the recoil velocity; conversely, future dynamical modelling of the BLRs around recoiling SMBHs, which also take into account the mass loss at the time of the merger, could yield better recoil-velocity estimates, and sharpen the test we discuss in this paper.

\subsection{Additional lines}
\label{ssec:other_lines}

We perform the same correlation analysis as for the H$\beta_{\rm br}$ line above, but estimating our redshift of the BLR, $z_{\rm B}$, using one of the three narrow lines ([O III] 5007, Ca II 3934, and [OII] 3728) listed in Section~\ref{ssec:Dv}.  Given that this effectively means comparing narrow lines with narrow lines, we do not expect any substantial correlation with $\Delta (g-i)$ under our hypothesis, so this serves as a null test of our results.

\begin{table*}
	\centering
	\caption{Spearman correlation test results treating various lines as the broad line. Note that H$\beta_{\rm br}$ is our main result, as this is a reliable broad line, while [O III] 5007, [O II] 3728 and Ca II 3934 are narrow lines, which are only listed here as null checks. We show the results for two variants of the analysis, with and without the broad line consistency cut discussed in Section~\ref{ssec:quality_cuts}.}
	\label{tab:corr_results}
	\begin{tabular}{ll|c|ccc} 
		\hline
		 & & H$\beta$ & [O III] 5007 & [O II] 3728 & Ca II 3934\\
		      \hline
		\multirow{2}{*}{Main analysis} & $r$ & 0.118 & -0.0028 & -0.0035 & -0.107\\
         & $p$-value & $\ll10^{-10}$ & 0.50 & 0.38 & $\ll10^{-10}$\\
                \hline
		\multirow{2}{*}{Without BLR consistency cut} & $r$ & 0.138 & -0.005 & -0.026 & -0.081\\
         & $p$-value & $\ll10^{-10}$ & 0.17 & $\ll10^{-10}$ & $\ll10^{-10}$\\
		\hline
	\end{tabular}
\end{table*}

\begin{figure*}
    \centering

    \begin{subfigure}{0.49\textwidth}
        \centering
        \includegraphics[width=\linewidth]{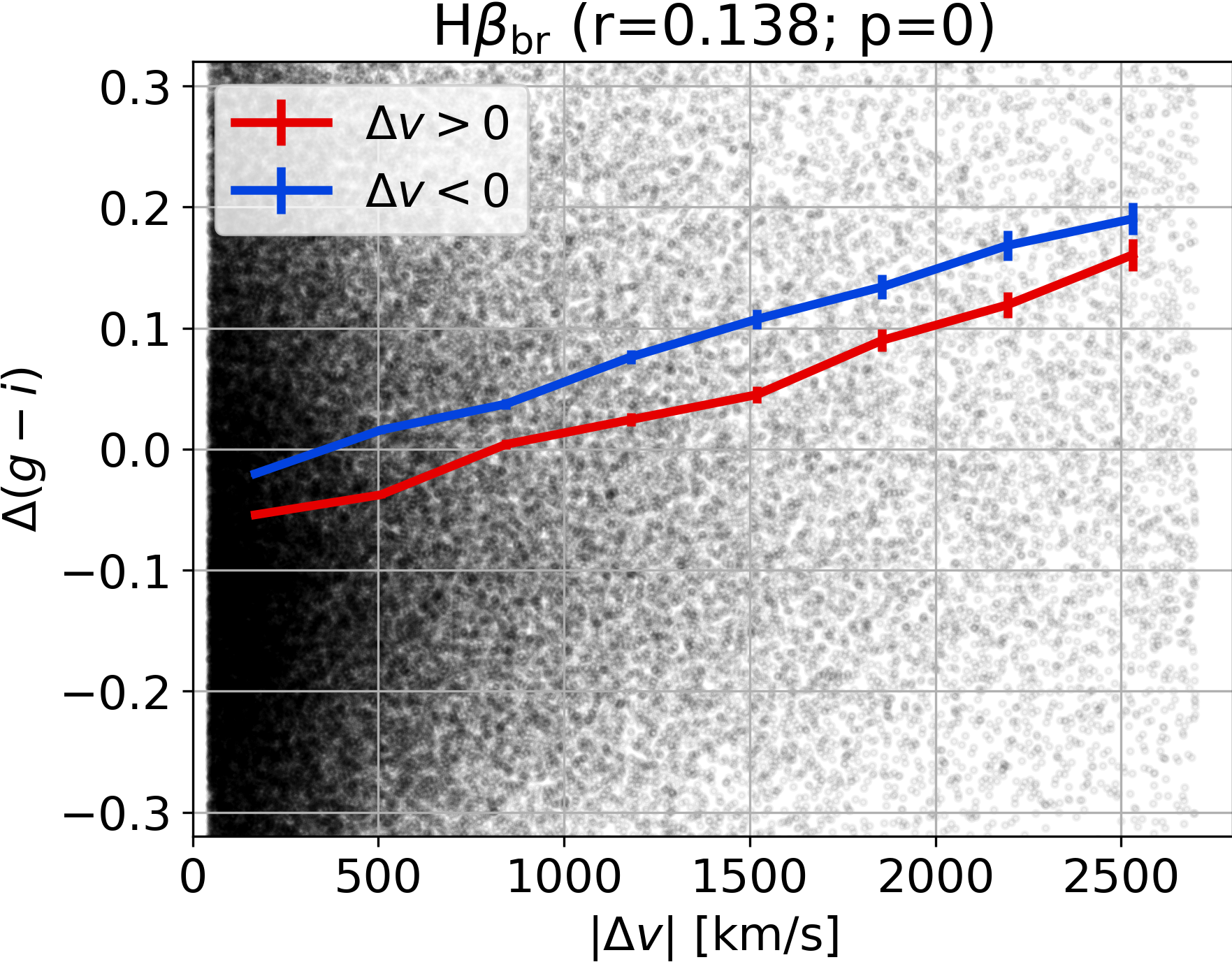}
    \end{subfigure}
    \hspace{0.01\textwidth}
    \begin{subfigure}{0.49\textwidth}
        \centering
        \includegraphics[width=\linewidth]{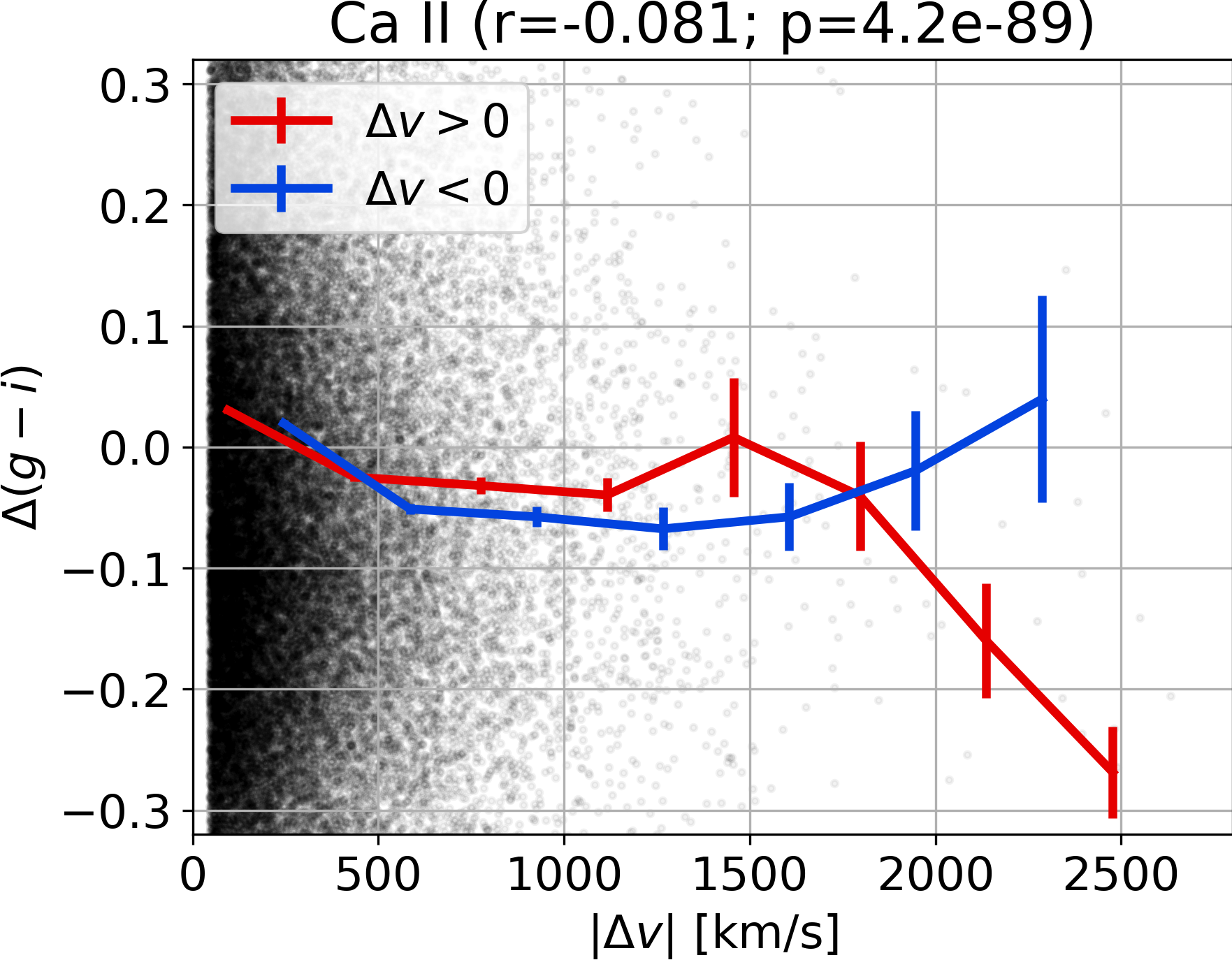}
    \end{subfigure}

    \vspace{0.5cm} 

    \begin{subfigure}{0.49\textwidth}
        \centering
        \includegraphics[width=\linewidth]{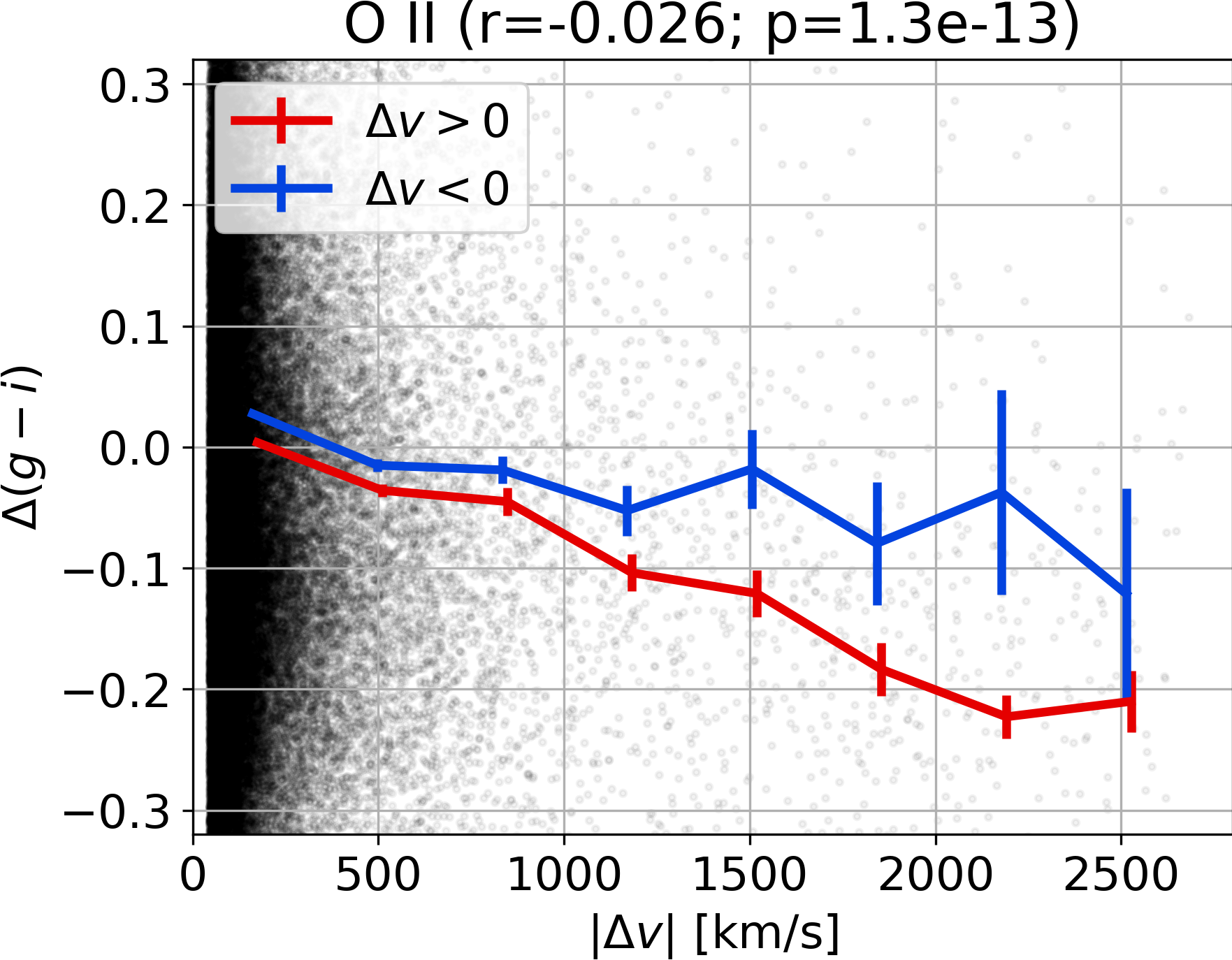}
    \end{subfigure}
    \hspace{0.01\textwidth}
    \begin{subfigure}{0.49\textwidth}
        \centering
        \includegraphics[width=\linewidth]{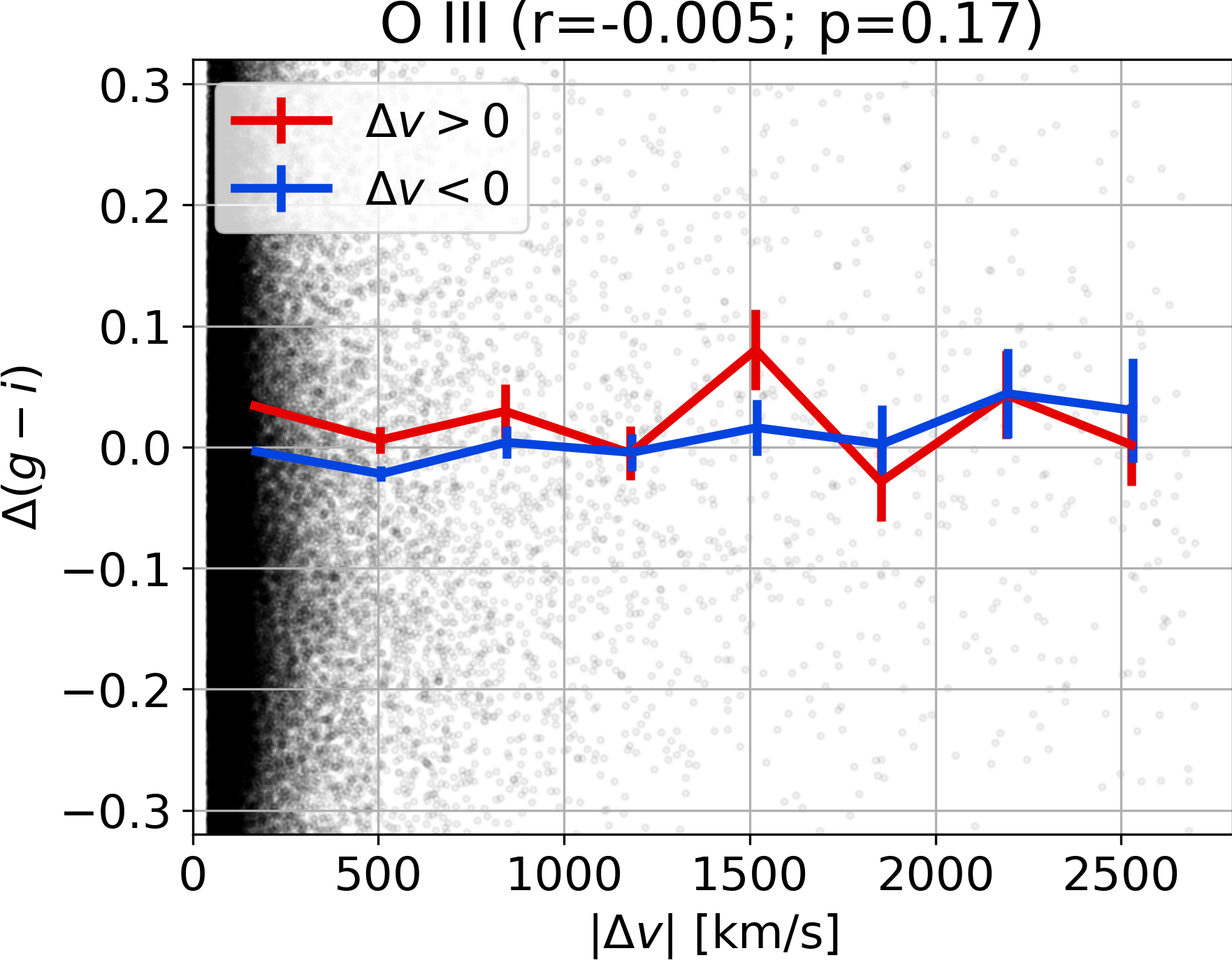}
    \end{subfigure}

    \caption{$\Delta (g-i)$ relative colour as a function of the absolute value of the line-of-sight peculiar velocity of the BLR, $|\Delta v|$, obtained by treating different lines as the broad line: H$\beta_{\rm br}$ (top left), Ca II 3934 (top right), [O II] 3728 (bottom left), [O III] 5007 (bottom right). Notation is the same as in Fig.~\ref{fig:dv_vs_dgmr_shift}, but we used the data without the BLR consistency cut, hence the difference between the top left panel and Fig.~\ref{fig:dv_vs_dgmr_shift}. This is necessary, because Ca II, [O II] and [O III] are narrow lines, so this cut would filter out a large fraction of our data. We can see that the H$\beta_{\rm br}$ results are relatively unchanged by the lack of BLR consistency cut. Treating narrow lines as broad lines serves as a consistency check, and we can see that they do not show such a clear correlation as our main result with H$\beta_{\rm br}$. No correlation is seen with [O III]. Ca II shows significance in the Spearman test, but no clear trend in the binned data. [O II] does show a significant anticorrelation, mostly by the redshifted subpopulation. This indicates that for some reason dust-obscured or intrinsically red QSOs are less likely to show inconsistent redshifts between different narrow lines, particularly [O II].}\label{fig:all_lines_wo_Haiman}
\end{figure*}

The number of QSOs for these various lines after different cuts are shown in Table~\ref{tab:qso_numbers}, and results of the Spearman correlation test are listed in Table~\ref{tab:corr_results}. We also show binned $\Delta v$ vs.~$\Delta (g-i)$ plots in Figure~\ref{fig:all_lines_wo_Haiman}. Note that here we are not applying the broad line consistency cut, since this is not appropriate for narrow lines and filters out a large portion of the data. We can see that the H$\beta_{\rm br}$ line shows even stronger correlation without this filter. We also find that none of the narrow lines show as clear a trend as H$\beta_{\rm br}$. However, they all show weak negative correlations with varied levels of significance. The correlation using [OIII] is insignificant, and the mean lines are visibly flat. The correlation when using Ca II is highly significant, however, the mean lines do not show a clear trend. They start out decreasing for both blue- and redshifted subpopulations, and later diverge. The high significance of the Spearman test is due to low velocity QSOs ($|\Delta v|\lesssim$ 1000 km/s), which dominate in number and show a clear trend in the mean lines. The correlation using [OII] is also significant, although, the correlation coefficient is itself quite small ($r=-0.026$). The mean lines also show a clear trend in this case, particularly for the redshifted subsample. 

One way to interpret these null tests is by recognising that when we treat one of the three narrow lines as a broad line, $\Delta v$ effectively measures how consistently these narrow lines trace the same NLR. We can see that there are QSOs where different narrow lines are shifted from each other by hundreds or even thousands of km/s. This in itself indicates, that at least for a subset of QSOs these narrow lines can be unreliable as a reference to the rest frame of the host galaxy or the obscuring dust torus. The negative correlations with $\Delta (g-i)$ also indicate that different narrow lines tend to show more disagreement in their measured redshift for blue QSOs. Whether this can have some physical explanation (e.g.~more prominent NLR winds in more blue QSOs), or is due to some selection effects remains to be seen, and will need to be investigated further in future work. However, it is clear that the distinct positive correlation we see for H$\beta_{\rm br}$ does not show up when comparing narrow lines, which increases the confidence in our main result.

\section{Conclusions and Future Work}
\label{sec:Conclusions}

Recoiling SMBHs are direct tracers of past SMBH mergers, and while there are several candidates \citep{Komossa2008,Robinson2010,BlechaCID42_2013,NovakCID42_2015,Chiaberge2017,Kalfountzou2017,Jadhav2021,Hogg2021,Barrows2025}, we do not yet have a way to rule out all alternative explanations. In this paper, we take a different approach, and perform the first search for a population-level signature of these recoiling SMBHs. We specifically look for a positive correlation between QSO dust reddening and the line-of-sight velocity of the BLR associated with the SMBH relative to the host galaxy. Such a correlation was predicted by simulations averaging over the orientation of the AGN disc, the recoil direction, and the time of observation after the recoil \citep{AGNOsc_theory2016}. It arises physically because a recoiling SMBH's velocity is largest when it is closest to the centre, where the line of sight more frequently intersects larger dust columns.

We analyse about 100,000 QSOs from the SDSS-DR16 QSO catalogue \citep{SDSS-DR16Q}, which have the required emission lines in band and pass various quality cuts (see Sec.~\ref{ssec:quality_cuts}). We use the $\Delta (g-i)$ relative colour as a proxy for the dust column density. The $v_{\rm SMBH}$ line-of-sight velocity of the SMBH relative to the central engine is approximated with the line-of-sight velocity of the BLR w.r.t~the NLR ($\Delta v$), which can be estimated from the difference between the redshift of the broad component of the H$\beta$ emission line, and a noise-weighted average redshift of narrow lines ([O III] 5007, [O II] 3728, and Ca II 3934)). We use spectral fits of SDSS QSOs from \citet{Wu_Shen2022}. We find a highly significant positive correlation between $|\Delta v|$ and $\Delta (g-i)$, which is consistent with the expected behaviour if a significant fraction of the sample represent recoiling SMBHs.

While the correlation we find might be evidence for a population of recoiling SMBHs, alternative explanations are possible. Both broad and narrow lines can be shifted from systemic due to various effects \citep{Shen2016_velocity_shifts}, so non-zero $\Delta v$ values can arise even in the absence of a genuine bulk motion of the BLR. It is not clear however why line shifts due to other effects would be correlated with dust reddening. Nevertheless, we carry out several consistency checks of our results. We find that the correlation is driven by truly dust-obscured QSOs, and not those having an intrinsically steep spectrum (see Fig.~\ref{fig:red_qso_fraction}). We find the correlation result to be robust against the lower limit we apply on $|\Delta v|$ in order to exclude the majority of QSOs, which are unrelated to mergers and recoils (see Fig.~\ref{fig:delta_v_limits}). The results also remain qualitatively unchanged over different ways to estimate line redshifts (see Fig.~\ref{fig:dv_sys_vs_peak_vs_cent}). We also see that most of our QSO sample naturally obeys the limit on recoil velocities and line widths needed for the SMBH to retain its BLR (see Fig.~\ref{fig:FWHM_vs_Dv}), which further supports our hypothesis that a large fraction of our sample represent recoiling SMBHs.

We find a difference in the dust obscuration of subpopulations where the BLR is moving towards or away from us (see Fig.~\ref{fig:dv_vs_dgmr_shift}). While some of this difference can be attributed to the specific way line redshifts are estimated (see Fig.~\ref{fig:dv_sys_vs_peak_vs_cent}), they cannot fully account for this effect. Future work could focus on simulations of the recoiling SMBH or alternative scenarios, to understand if such an asymmetry might be expected. Analysis using other broad emission lines would also be valuable to understand if the asymmetry is consistent across different lines.

We also find that the redshifts of the three narrow lines used ([O III] 5007, [O II] 3728, and Ca II 3934)) can be highly inconsistent in cases, resulting in hundreds or even thousands of km/s shifts between them (see Fig.~\ref{fig:all_lines_wo_Haiman}). These differences also tend to be more prominent for blue QSOs with low values of $\Delta (g-i)$. While this results in a negative correlation between $\Delta (g-i)$ and $|\Delta v|$ derived from narrow lines only, we believe that this likely cannot strongly influence our main results. This is because the correlations found in this case are weaker and less consistent than our main result with H$\beta_{\rm br}$, and they are driven by blue QSOs ($\Delta (g-i)\lesssim 0$), not dust-reddened ones ($\Delta (g-i)\gtrsim 0.2-0.4$). However, a more extensive investigation would be beneficial in the future.

This work represents the first population-level observational search for recoiling SMBHs. If the observed correlation is indeed driven by recoil, it opens up the possibility of constraining both the SMBH merger population and the environments in which these mergers occur. Several avenues for follow-up will be important to test and refine this interpretation. Extending the analysis to other broad emission lines would help establish the robustness of the signal, while estimating dust reddening directly from fitted spectra, rather than photometry, could reduce systematic uncertainties. For example, \citet{Dong2008} show that the broad H$\alpha$/H$\beta$ ratio in a large homogeneous AGN sample peaks at $\sim$3.06 with a small dispersion, suggesting that it can serve as a statistical indicator of BLR reddening. Beyond correlation-based approaches, a full Bayesian analysis of the QSO population would provide a more rigorous framework for inference, and would enable direct constraints on SMBH population properties, recoil velocities, and AGN disc parameters.

Assuming that the measured correlation is indeed caused by recoiling SMBHs, it is already possible to draw preliminary conclusions about the fraction of QSOs that have undergone a recent SMBH merger. \citet{AGNOsc_theory2016} report a Pearson correlation coefficient of $r=0.28$ under the assumption that all QSOs are recoiling following a recent merger. Comparing this to our measured value of $r=0.13$ suggests that approximately $\sim 50\%$ of QSOs may have experienced a recent merger. Interpreting this fraction as a duty cycle ($f$), the SMBH merger rate per QSO can be estimated as $R \sim f/t_{\rm QSO}$, where $t_{\rm QSO} \sim 10^7-10^8\,{\rm yr}$ is the expected duration of the luminous QSO phase of a recoiling SMBH \citep{EjectedQSO_timescale_Loeb2007}. This yields an approximate merger rate of $R \approx 5\times(10^{-9}-10^{-8})\,{\rm yr}^{-1}$ per SMBH.

This estimate is, however, subject to significant uncertainties. It depends sensitively on the assumptions underlying the simulations of \citet{AGNOsc_theory2016}, as well as on the adopted visibility timescale of the recoiling phase. In particular, uncertainties in how the observed correlation maps onto the fraction of post-merger systems, and in how long such systems remain observable, propagate directly into order-of-magnitude uncertainties in the inferred merger rate. Despite this, the presence of the correlation itself, if indeed driven by recoiling SMBHs, robustly implies that a substantial fraction of QSOs must have experienced a recent SMBH merger. If confirmed, this would have important implications for the expected merger rate of $\sim10^5$–$10^7\,M_\odot$ black holes, suggesting that LISA may observe a correspondingly high rate of events associated with QSO hosts. It would also enable a direct connection to existing searches for spatially and spectroscopically offset BLRs (e.g.~\citealt{Barrows2025}), as the inferred post-merger fraction can be translated into a predicted incidence of offset QSOs; agreement between these predictions and observed samples would provide a powerful and independent test of the recoiling SMBH scenario.

\section*{Acknowledgements}
We thank Paul Hewett for useful discussions, and Qiaoya Wu for guidance on the data presented in \citet{Wu_Shen2022}.
ZH acknowledges financial support from NASA grants 80NSSC24K0440 and 80NSSC22K0822.
PR and ZF have received funding from the HUN-REN Hungarian Research Network and was supported by the NKFIH excellence grant TKP2021-NKTA-64.

\section*{Data Availability} 

The data and software that support the findings of this study are openly available. The processed datasets, analysis outputs, and software are archived on Zenodo \citep{zenodo_code_release}, and are also available on GitHub \citep{github_code_release}.



\bibliographystyle{mnras}
\bibliography{AGN_Oscillator} 

@software{zenodo_code_release,
  author       = {B\'ecsy, Bence and
                  Raffai, Peter and
                  Haiman, Zoltan and
                  Budai, Andor and
                  Frei, Zsolt},
  title        = {bencebecsy/AGNOsc: v1.0.0},
  month        = may,
  year         = 2026,
  publisher    = {Zenodo},
  version      = {v1.0.0},
  doi          = {10.5281/zenodo.20041006},
}

@software{github_code_release,
  author       = {{B\'ecsy}, B. and {Raffai}, P. and {Haiman}, Z. and {Budai}, A. and {Frei}, Z.},
  title        = {AGNOsc},
  year         = {2026},
  url          = {https://github.com/bencebecsy/AGNOsc},
  note         = {Accessed: 2026-05-05}
}

@ARTICLE{Shen2019,
       author = {{Shen}, Yue and {Hall}, Patrick B. and {Horne}, Keith and {Zhu}, Guangtun and {McGreer}, Ian and {Simm}, Torben and {Trump}, Jonathan R. and {Kinemuchi}, Karen and {Brandt}, W.~N. and {Green}, Paul J. and {Grier}, C.~J. and {Guo}, Hengxiao and {Ho}, Luis C. and {Homayouni}, Yasaman and {Jiang}, Linhua and {I-Hsiu Li}, Jennifer and {Morganson}, Eric and {Petitjean}, Patrick and {Richards}, Gordon T. and {Schneider}, Donald P. and {Starkey}, D.~A. and {Wang}, Shu and {Chambers}, Ken and {Kaiser}, Nick and {Kudritzki}, Rolf-Peter and {Magnier}, Eugene and {Waters}, Christopher},
        title = "{The Sloan Digital Sky Survey Reverberation Mapping Project: Sample Characterization}",
      journal = {\apjs},
         year = 2019,
        month = apr,
       volume = {241},
       number = {2},
          eid = {34},
        pages = {34},
          doi = {10.3847/1538-4365/ab074f},
archivePrefix = {arXiv},
       eprint = {1810.01447},
 primaryClass = {astro-ph.GA},
       adsurl = {https://ui.adsabs.harvard.edu/abs/2019ApJS..241...34S}
}

@ARTICLE{Shen2011_DR7,
       author = {{Shen}, Yue and {Richards}, Gordon T. and {Strauss}, Michael A. and {Hall}, Patrick B. and {Schneider}, Donald P. and {Snedden}, Stephanie and {Bizyaev}, Dmitry and {Brewington}, Howard and {Malanushenko}, Viktor and {Malanushenko}, Elena and {Oravetz}, Dan and {Pan}, Kaike and {Simmons}, Audrey},
        title = "{A Catalog of Quasar Properties from Sloan Digital Sky Survey Data Release 7}",
      journal = {\apjs},
         year = 2011,
        month = jun,
       volume = {194},
       number = {2},
          eid = {45},
        pages = {45},
          doi = {10.1088/0067-0049/194/2/45},
archivePrefix = {arXiv},
       eprint = {1006.5178},
 primaryClass = {astro-ph.CO},
       adsurl = {https://ui.adsabs.harvard.edu/abs/2011ApJS..194...45S}
}

@ARTICLE{virial_mass_Vestergaard2006,
       author = {{Vestergaard}, Marianne and {Peterson}, Bradley M.},
        title = "{Determining Central Black Hole Masses in Distant Active Galaxies and Quasars. II. Improved Optical and UV Scaling Relationships}",
      journal = {\apj},
         year = 2006,
        month = apr,
       volume = {641},
       number = {2},
        pages = {689-709},
          doi = {10.1086/500572},
archivePrefix = {arXiv},
       eprint = {astro-ph/0601303},
 primaryClass = {astro-ph},
       adsurl = {https://ui.adsabs.harvard.edu/abs/2006ApJ...641..689V}
}

@ARTICLE{Hogg2021,
       author = {{Hogg}, J. Drew and {Blecha}, Laura and {Reynolds}, Christopher S. and {Smith}, Krista Lynne and {Winter}, Lisa M.},
        title = "{2MASX J00423991 + 3017515: an offset active galactic nucleus in an interacting system}",
      journal = {\mnras},
         year = 2021,
        month = may,
       volume = {503},
       number = {2},
        pages = {1688-1702},
          doi = {10.1093/mnras/stab576},
archivePrefix = {arXiv},
       eprint = {2103.00012},
 primaryClass = {astro-ph.GA},
       adsurl = {https://ui.adsabs.harvard.edu/abs/2021MNRAS.503.1688H}
}

@ARTICLE{Li2024,
       author = {{Li}, Junyao and {Zhuang}, Ming-Yang and {Shen}, Yue},
        title = "{JWST Confirms the Nature of CID-42}",
      journal = {\apj},
         year = 2024,
        month = jan,
       volume = {961},
       number = {1},
          eid = {19},
        pages = {19},
          doi = {10.3847/1538-4357/ad0e0d},
archivePrefix = {arXiv},
       eprint = {2307.05852},
 primaryClass = {astro-ph.GA},
       adsurl = {https://ui.adsabs.harvard.edu/abs/2024ApJ...961...19L}
}

@ARTICLE{Kalfountzou2017,
       author = {{Kalfountzou}, E. and {Santos Lleo}, M. and {Trichas}, M.},
        title = "{SDSS J1056+5516: A Triple AGN or an SMBH Recoil Candidate?}",
      journal = {\apjl},
         year = 2017,
        month = dec,
       volume = {851},
       number = {1},
          eid = {L15},
        pages = {L15},
          doi = {10.3847/2041-8213/aa9b2d},
archivePrefix = {arXiv},
       eprint = {1712.03909},
 primaryClass = {astro-ph.GA},
       adsurl = {https://ui.adsabs.harvard.edu/abs/2017ApJ...851L..15K}
}

@ARTICLE{Komossa2008,
       author = {{Komossa}, S. and {Zhou}, H. and {Lu}, H.},
        title = "{A Recoiling Supermassive Black Hole in the Quasar SDSS J092712.65+294344.0?}",
      journal = {\apjl},
         year = 2008,
        month = may,
       volume = {678},
       number = {2},
        pages = {L81},
          doi = {10.1086/588656},
archivePrefix = {arXiv},
       eprint = {0804.4585},
 primaryClass = {astro-ph},
       adsurl = {https://ui.adsabs.harvard.edu/abs/2008ApJ...678L..81K}
}

@ARTICLE{LISA_SMBHB_Klein2016,
       author = {{Klein}, Antoine and {Barausse}, Enrico and {Sesana}, Alberto and {Petiteau}, Antoine and {Berti}, Emanuele and {Babak}, Stanislav and {Gair}, Jonathan and {Aoudia}, Sofiane and {Hinder}, Ian and {Ohme}, Frank and {Wardell}, Barry},
        title = "{Science with the space-based interferometer eLISA: Supermassive black hole binaries}",
      journal = {\prd},
         year = 2016,
        month = jan,
       volume = {93},
       number = {2},
          eid = {024003},
        pages = {024003},
          doi = {10.1103/PhysRevD.93.024003},
archivePrefix = {arXiv},
       eprint = {1511.05581},
 primaryClass = {gr-qc},
       adsurl = {https://ui.adsabs.harvard.edu/abs/2016PhRvD..93b4003K}
}

@ARTICLE{LISAAmaro-Seoane2017,
       author = {{Amaro-Seoane}, Pau and {Audley}, Heather and {Babak}, Stanislav and {Baker}, John and {Barausse}, Enrico and {Bender}, Peter and {Berti}, Emanuele and {Binetruy}, Pierre and {Born}, Michael and {Bortoluzzi}, Daniele and {Camp}, Jordan and {Caprini}, Chiara and {Cardoso}, Vitor and {Colpi}, Monica and {Conklin}, John and {Cornish}, Neil and {Cutler}, Curt and {Danzmann}, Karsten and {Dolesi}, Rita and {Ferraioli}, Luigi and {Ferroni}, Valerio and {Fitzsimons}, Ewan and {Gair}, Jonathan and {Gesa Bote}, Lluis and {Giardini}, Domenico and {Gibert}, Ferran and {Grimani}, Catia and {Halloin}, Hubert and {Heinzel}, Gerhard and {Hertog}, Thomas and {Hewitson}, Martin and {Holley-Bockelmann}, Kelly and {Hollington}, Daniel and {Hueller}, Mauro and {Inchauspe}, Henri and {Jetzer}, Philippe and {Karnesis}, Nikos and {Killow}, Christian and {Klein}, Antoine and {Klipstein}, Bill and {Korsakova}, Natalia and {Larson}, Shane L and {Livas}, Jeffrey and {Lloro}, Ivan and {Man}, Nary and {Mance}, Davor and {Martino}, Joseph and {Mateos}, Ignacio and {McKenzie}, Kirk and {McWilliams}, Sean T and {Miller}, Cole and {Mueller}, Guido and {Nardini}, Germano and {Nelemans}, Gijs and {Nofrarias}, Miquel and {Petiteau}, Antoine and {Pivato}, Paolo and {Plagnol}, Eric and {Porter}, Ed and {Reiche}, Jens and {Robertson}, David and {Robertson}, Norna and {Rossi}, Elena and {Russano}, Giuliana and {Schutz}, Bernard and {Sesana}, Alberto and {Shoemaker}, David and {Slutsky}, Jacob and {Sopuerta}, Carlos F. and {Sumner}, Tim and {Tamanini}, Nicola and {Thorpe}, Ira and {Troebs}, Michael and {Vallisneri}, Michele and {Vecchio}, Alberto and {Vetrugno}, Daniele and {Vitale}, Stefano and {Volonteri}, Marta and {Wanner}, Gudrun and {Ward}, Harry and {Wass}, Peter and {Weber}, William and {Ziemer}, John and {Zweifel}, Peter},
        title = "{Laser Interferometer Space Antenna}",
      journal = {arXiv e-prints},
         year = 2017,
        month = feb,
          eid = {arXiv:1702.00786},
        pages = {arXiv:1702.00786},
          doi = {10.48550/arXiv.1702.00786},
archivePrefix = {arXiv},
       eprint = {1702.00786},
 primaryClass = {astro-ph.IM},
       adsurl = {https://ui.adsabs.harvard.edu/abs/2017arXiv170200786A}
}

@ARTICLE{NovakCID42_2015,
       author = {{Novak}, Mladen and {Smol{\v{c}}i{\'c}}, Vernesa and {Civano}, Francesca and {Bondi}, Marco and {Ciliegi}, Paolo and {Wang}, Xiawei and {Loeb}, Abraham and {Banfield}, Julie and {Bourke}, Stephen and {Elvis}, Martin and {Hallinan}, Gregg and {Intema}, Huib T. and {Kl{\"o}ckner}, Hans-Rainer and {Mooley}, Kunal and {Navarrete}, Felipe},
        title = "{New insights from deep VLA data on the potentially recoiling black hole CID-42 in the COSMOS field}",
      journal = {\mnras},
         year = 2015,
        month = feb,
       volume = {447},
       number = {2},
        pages = {1282-1288},
          doi = {10.1093/mnras/stu2473},
archivePrefix = {arXiv},
       eprint = {1412.0004},
 primaryClass = {astro-ph.GA},
       adsurl = {https://ui.adsabs.harvard.edu/abs/2015MNRAS.447.1282N}
}

@ARTICLE{BlechaCID42_2013,
       author = {{Blecha}, Laura and {Civano}, Francesca and {Elvis}, Martin and {Loeb}, Abraham},
        title = "{Constraints on the nature of CID-42: recoil kick or supermassive black hole pair?}",
      journal = {\mnras},
         year = 2013,
        month = jan,
       volume = {428},
       number = {2},
        pages = {1341-1350},
          doi = {10.1093/mnras/sts114},
archivePrefix = {arXiv},
       eprint = {1205.6202},
 primaryClass = {astro-ph.CO},
       adsurl = {https://ui.adsabs.harvard.edu/abs/2013MNRAS.428.1341B}
}

@ARTICLE{LippaiFreiHaiman2008,
       author = {{Lippai}, Zolt{\'a}n and {Frei}, Zsolt and {Haiman}, Zolt{\'a}n},
        title = "{Prompt Shocks in the Gas Disk around a Recoiling Supermassive Black Hole Binary}",
      journal = {\apjl},
         year = 2008,
        month = mar,
       volume = {676},
       number = {1},
        pages = {L5},
          doi = {10.1086/587034},
archivePrefix = {arXiv},
       eprint = {0801.0739},
 primaryClass = {astro-ph},
       adsurl = {https://ui.adsabs.harvard.edu/abs/2008ApJ...676L...5L}
}

@ARTICLE{HewettWildQSOz2010,
       author = {{Hewett}, Paul C. and {Wild}, Vivienne},
        title = "{Improved redshifts for SDSS quasar spectra}",
      journal = {\mnras},
         year = 2010,
        month = jul,
       volume = {405},
       number = {4},
        pages = {2302-2316},
          doi = {10.1111/j.1365-2966.2010.16648.x},
archivePrefix = {arXiv},
       eprint = {1003.3017},
 primaryClass = {astro-ph.CO},
       adsurl = {https://ui.adsabs.harvard.edu/abs/2010MNRAS.405.2302H}
}

@BOOK{KrolikAGNBook1999,
       author = {{Krolik}, Julian H.},
        title = "{Active Galactic Nuclei. From the Central Black Hole to the Galactic Environment}",
         year = 1999,
       adsurl = {https://ui.adsabs.harvard.edu/abs/1999agnf.book.....K}
}

@ARTICLE{Bentz2013HBeta,
       author = {{Bentz}, Misty C. and {Denney}, Kelly D. and {Grier}, Catherine J. and {Barth}, Aaron J. and {Peterson}, Bradley M. and {Vestergaard}, Marianne and {Bennert}, Vardha N. and {Canalizo}, Gabriela and {De Rosa}, Gisella and {Filippenko}, Alexei V. and {Gates}, Elinor L. and {Greene}, Jenny E. and {Li}, Weidong and {Malkan}, Matthew A. and {Pogge}, Richard W. and {Stern}, Daniel and {Treu}, Tommaso and {Woo}, Jong-Hak},
        title = "{The Low-luminosity End of the Radius-Luminosity Relationship for Active Galactic Nuclei}",
      journal = {\apj},
         year = 2013,
        month = apr,
       volume = {767},
       number = {2},
          eid = {149},
        pages = {149},
          doi = {10.1088/0004-637X/767/2/149},
archivePrefix = {arXiv},
       eprint = {1303.1742},
 primaryClass = {astro-ph.CO},
       adsurl = {https://ui.adsabs.harvard.edu/abs/2013ApJ...767..149B}
}

@ARTICLE{Bogdanociv_et_al_LRR2022,
       author = {{Bogdanovi{\'c}}, Tamara and {Miller}, M. Coleman and {Blecha}, Laura},
        title = "{Electromagnetic counterparts to massive black-hole mergers}",
      journal = {Living Reviews in Relativity},
         year = 2022,
        month = dec,
       volume = {25},
       number = {1},
          eid = {3},
        pages = {3},
          doi = {10.1007/s41114-022-00037-8},
archivePrefix = {arXiv},
       eprint = {2109.03262},
 primaryClass = {astro-ph.HE},
       adsurl = {https://ui.adsabs.harvard.edu/abs/2022LRR....25....3B}
}

@ARTICLE{Dorazio_Charisi_EMSigReview2023,
       author = {{D'Orazio}, Daniel J. and {Charisi}, Maria},
        title = "{Observational Signatures of Supermassive Black Hole Binaries}",
      journal = {arXiv e-prints},
         year = 2023,
        month = oct,
          eid = {arXiv:2310.16896},
        pages = {arXiv:2310.16896},
          doi = {10.48550/arXiv.2310.16896},
archivePrefix = {arXiv},
       eprint = {2310.16896},
 primaryClass = {astro-ph.HE},
       adsurl = {https://ui.adsabs.harvard.edu/abs/2023arXiv231016896D}
}

@ARTICLE{KormendyHo2013,
       author = {{Kormendy}, John and {Ho}, Luis C.},
        title = "{Coevolution (Or Not) of Supermassive Black Holes and Host Galaxies}",
      journal = {\araa},
         year = 2013,
        month = aug,
       volume = {51},
       number = {1},
        pages = {511-653},
          doi = {10.1146/annurev-astro-082708-101811},
archivePrefix = {arXiv},
       eprint = {1304.7762},
 primaryClass = {astro-ph.CO},
       adsurl = {https://ui.adsabs.harvard.edu/abs/2013ARA&A..51..511K}
}

@ARTICLE{SMBHB_SDSS_Eracleous2012,
       author = {{Eracleous}, Michael and {Boroson}, Todd A. and {Halpern}, Jules P. and {Liu}, Jia},
        title = "{A Large Systematic Search for Close Supermassive Binary and Rapidly Recoiling Black Holes}",
      journal = {\apjs},
         year = 2012,
        month = aug,
       volume = {201},
       number = {2},
          eid = {23},
        pages = {23},
          doi = {10.1088/0067-0049/201/2/23},
archivePrefix = {arXiv},
       eprint = {1106.2952},
 primaryClass = {astro-ph.CO},
       adsurl = {https://ui.adsabs.harvard.edu/abs/2012ApJS..201...23E}
}

@ARTICLE{SMBHB_SDSS_Ju2013,
       author = {{Ju}, Wenhua and {Greene}, Jenny E. and {Rafikov}, Roman R. and {Bickerton}, Steven J. and {Badenes}, Carles},
        title = "{Search for Supermassive Black Hole Binaries in the Sloan Digital Sky Survey Spectroscopic Sample}",
      journal = {\apj},
         year = 2013,
        month = nov,
       volume = {777},
       number = {1},
          eid = {44},
        pages = {44},
          doi = {10.1088/0004-637X/777/1/44},
archivePrefix = {arXiv},
       eprint = {1306.4987},
 primaryClass = {astro-ph.CO},
       adsurl = {https://ui.adsabs.harvard.edu/abs/2013ApJ...777...44J}
}

@ARTICLE{offset_BLR_SMBHBs_Liu2014,
       author = {{Liu}, Xin and {Shen}, Yue and {Bian}, Fuyan and {Loeb}, Abraham and {Tremaine}, Scott},
        title = "{Constraining Sub-parsec Binary Supermassive Black Holes in Quasars with Multi-epoch Spectroscopy. II. The Population with Kinematically Offset Broad Balmer Emission Lines}",
      journal = {\apj},
         year = 2014,
        month = jul,
       volume = {789},
       number = {2},
          eid = {140},
        pages = {140},
          doi = {10.1088/0004-637X/789/2/140},
archivePrefix = {arXiv},
       eprint = {1312.6694},
 primaryClass = {astro-ph.CO},
       adsurl = {https://ui.adsabs.harvard.edu/abs/2014ApJ...789..140L}
}

@ARTICLE{offaxis_var_Gaskell2011,
       author = {{Gaskell}, C. Martin},
        title = "{Off-Axis Variability of AGNs: a New Paradigm for Broad Lines and Continuum Emitting Regions}",
      journal = {Baltic Astronomy},
         year = 2011,
        month = aug,
       volume = {20},
        pages = {392-399},
          doi = {10.1515/astro-2017-0309},
archivePrefix = {arXiv},
       eprint = {1107.5382},
 primaryClass = {astro-ph.CO},
       adsurl = {https://ui.adsabs.harvard.edu/abs/2011BaltA..20..392G}
}

@ARTICLE{disk_Chen1989,
       author = {{Chen}, Kaiyou and {Halpern}, Jules P. and {Filippenko}, Alexei V.},
        title = "{Kinematic Evidence for a Relativistic Keplerian Disk: ARP 102B}",
      journal = {\apj},
         year = 1989,
        month = apr,
       volume = {339},
        pages = {742},
          doi = {10.1086/167332},
       adsurl = {https://ui.adsabs.harvard.edu/abs/1989ApJ...339..742C}
}

@ARTICLE{inflow_Gaskell2016,
       author = {{Gaskell}, C. Martin and {Goosmann}, Ren{\'e} W.},
        title = "{The case for inflow of the broad-line region of active galactic nuclei}",
      journal = {\apss},
         year = 2016,
        month = feb,
       volume = {361},
          eid = {67},
        pages = {67},
          doi = {10.1007/s10509-015-2648-1},
archivePrefix = {arXiv},
       eprint = {1512.08900},
 primaryClass = {astro-ph.GA},
       adsurl = {https://ui.adsabs.harvard.edu/abs/2016Ap&SS.361...67G}
}

@ARTICLE{Gaskell1982,
       author = {{Gaskell}, C.~M.},
        title = "{A redshift difference between high and low ionization emission-line regions in QSO's-evidence for radial motions.}",
      journal = {\apj},
         year = 1982,
        month = dec,
       volume = {263},
        pages = {79-86},
          doi = {10.1086/160481},
       adsurl = {https://ui.adsabs.harvard.edu/abs/1982ApJ...263...79G}
}

@ARTICLE{NLR_inclination_Fischer2013,
       author = {{Fischer}, T.~C. and {Crenshaw}, D.~M. and {Kraemer}, S.~B. and {Schmitt}, H.~R.},
        title = "{Determining Inclinations of Active Galactic Nuclei via their Narrow-line Region Kinematics. I. Observational Results}",
      journal = {\apjs},
         year = 2013,
        month = nov,
       volume = {209},
       number = {1},
          eid = {1},
        pages = {1},
          doi = {10.1088/0067-0049/209/1/1},
archivePrefix = {arXiv},
       eprint = {1308.4129},
 primaryClass = {astro-ph.CO},
       adsurl = {https://ui.adsabs.harvard.edu/abs/2013ApJS..209....1F}
}

@ARTICLE{NRL_offset_Crenshaw2010,
       author = {{Crenshaw}, D.~M. and {Schmitt}, H.~R. and {Kraemer}, S.~B. and {Mushotzky}, R.~F. and {Dunn}, J.~P.},
        title = "{Radial Velocity Offsets Due to Mass Outflows and Extinction in Active Galactic Nuclei}",
      journal = {\apj},
         year = 2010,
        month = jan,
       volume = {708},
       number = {1},
        pages = {419-426},
          doi = {10.1088/0004-637X/708/1/419},
archivePrefix = {arXiv},
       eprint = {0911.0675},
 primaryClass = {astro-ph.CO},
       adsurl = {https://ui.adsabs.harvard.edu/abs/2010ApJ...708..419C}
}

@ARTICLE{CIV_wind_Richards2011,
       author = {{Richards}, Gordon T. and {Kruczek}, Nicholas E. and {Gallagher}, S.~C. and {Hall}, Patrick B. and {Hewett}, Paul C. and {Leighly}, Karen M. and {Deo}, Rajesh P. and {Kratzer}, Rachael M. and {Shen}, Yue},
        title = "{Unification of Luminous Type 1 Quasars through C IV Emission}",
      journal = {\aj},
         year = 2011,
        month = may,
       volume = {141},
       number = {5},
          eid = {167},
        pages = {167},
          doi = {10.1088/0004-6256/141/5/167},
archivePrefix = {arXiv},
       eprint = {1011.2282},
 primaryClass = {astro-ph.GA},
       adsurl = {https://ui.adsabs.harvard.edu/abs/2011AJ....141..167R}
}

@ARTICLE{wind_emission_lines_Murray1995,
       author = {{Murray}, N. and {Chiang}, J. and {Grossman}, S.~A. and {Voit}, G.~M.},
        title = "{Accretion Disk Winds from Active Galactic Nuclei}",
      journal = {\apj},
         year = 1995,
        month = oct,
       volume = {451},
        pages = {498},
          doi = {10.1086/176238},
       adsurl = {https://ui.adsabs.harvard.edu/abs/1995ApJ...451..498M}
}

@ARTICLE{epta_dr2_interp,
       author = {{EPTA Collaboration} and {InPTA Collaboration} and {Antoniadis}, J. and {Arumugam}, P. and {Arumugam}, S. and {Babak}, S. and {Bagchi}, M. and {Bak Nielsen}, A.-S. and {Bassa}, C.~G. and {Bathula}, A. and {Berthereau}, A. and {Bonetti}, M. and {Bortolas}, E. and {Brook}, P.~R. and {Burgay}, M. and {Caballero}, R.~N. and {Chalumeau}, A. and {Champion}, D.~J. and {Chanlaridis}, S. and {Chen}, S. and {Cognard}, I. and {Dandapat}, S. and {Deb}, D. and {Desai}, S. and {Desvignes}, G. and {Dhanda-Batra}, N. and {Dwivedi}, C. and {Falxa}, M. and {Ferdman}, R.~D. and {Franchini}, A. and {Gair}, J.~R. and {Goncharov}, B. and {Gopakumar}, A. and {Graikou}, E. and {Grie{\ss}meier}, J.-M. and {Gualandris}, A. and {Guillemot}, L. and {Guo}, Y.~J. and {Gupta}, Y. and {Hisano}, S. and {Hu}, H. and {Iraci}, F. and {Izquierdo-Villalba}, D. and {Jang}, J. and {Jawor}, J. and {Janssen}, G.~H. and {Jessner}, A. and {Joshi}, B.~C. and {Kareem}, F. and {Karuppusamy}, R. and {Keane}, E.~F. and {Keith}, M.~J. and {Kharbanda}, D. and {Kikunaga}, T. and {Kolhe}, N. and {Kramer}, M. and {Krishnakumar}, M.~A. and {Lackeos}, K. and {Lee}, K.~J. and {Liu}, K. and {Liu}, Y. and {Lyne}, A.~G. and {McKee}, J.~W. and {Maan}, Y. and {Main}, R.~A. and {Mickaliger}, M.~B. and {Ni{\c{t}}u}, I.~C. and {Nobleson}, K. and {Paladi}, A.~K. and {Parthasarathy}, A. and {Perera}, B.~B.~P. and {Perrodin}, D. and {Petiteau}, A. and {Porayko}, N.~K. and {Possenti}, A. and {Prabu}, T. and {Quelquejay Leclere}, H. and {Rana}, P. and {Samajdar}, A. and {Sanidas}, S.~A. and {Sesana}, A. and {Shaifullah}, G. and {Singha}, J. and {Speri}, L. and {Spiewak}, R. and {Srivastava}, A. and {Stappers}, B.~W. and {Surnis}, M. and {Susarla}, S.~C. and {Susobhanan}, A. and {Takahashi}, K. and {Tarafdar}, P. and {Theureau}, G. and {Tiburzi}, C. and {van der Wateren}, E. and {Vecchio}, A. and {Venkatraman Krishnan}, V. and {Verbiest}, J.~P.~W. and {Wang}, J. and {Wang}, L. and {Wu}, Z. and {Auclair}, P. and {Barausse}, E. and {Caprini}, C. and {Crisostomi}, M. and {Fastidio}, F. and {Khizriev}, T. and {Middleton}, H. and {Neronov}, A. and {Postnov}, K. and {Roper Pol}, A. and {Semikoz}, D. and {Smarra}, C. and {Steer}, D.~A. and {Truant}, R.~J. and {Valtolina}, S.},
        title = "{The second data release from the European Pulsar Timing Array. IV. Implications for massive black holes, dark matter, and the early Universe}",
      journal = {\aap},
         year = 2024,
        month = may,
       volume = {685},
          eid = {A94},
        pages = {A94},
          doi = {10.1051/0004-6361/202347433},
archivePrefix = {arXiv},
       eprint = {2306.16227},
 primaryClass = {astro-ph.CO},
       adsurl = {https://ui.adsabs.harvard.edu/abs/2024A&A...685A..94E}
}

@ARTICLE{dgmiRichards2003,
       author = {{Richards}, Gordon T. and {Hall}, Patrick B. and {Vanden Berk}, Daniel E. and {Strauss}, Michael A. and {Schneider}, Donald P. and {Weinstein}, Michael A. and {Reichard}, Timothy A. and {York}, Donald G. and {Knapp}, G.~R. and {Fan}, Xiaohui and {Ivezi{\'c}}, {\v{Z}}eljko and {Brinkmann}, J. and {Budav{\'a}ri}, Tam{\'a}s and {Csabai}, Istv{\'a}n and {Nichol}, R.~C.},
        title = "{Red and Reddened Quasars in the Sloan Digital Sky Survey}",
      journal = {\aj},
         year = 2003,
        month = sep,
       volume = {126},
       number = {3},
        pages = {1131-1147},
          doi = {10.1086/377014},
archivePrefix = {arXiv},
       eprint = {astro-ph/0305305},
 primaryClass = {astro-ph},
       adsurl = {https://ui.adsabs.harvard.edu/abs/2003AJ....126.1131R}
}

@ARTICLE{Shen2016_velocity_shifts,
       author = {{Shen}, Yue and {Brandt}, W.~N. and {Richards}, Gordon T. and {Denney}, Kelly D. and {Greene}, Jenny E. and {Grier}, C.~J. and {Ho}, Luis C. and {Peterson}, Bradley M. and {Petitjean}, Patrick and {Schneider}, Donald P. and {Tao}, Charling and {Trump}, Jonathan R.},
        title = "{The Sloan Digital Sky Survey Reverberation Mapping Project: Velocity Shifts of Quasar Emission Lines}",
      journal = {\apj},
         year = 2016,
        month = nov,
       volume = {831},
       number = {1},
          eid = {7},
        pages = {7},
          doi = {10.3847/0004-637X/831/1/7},
archivePrefix = {arXiv},
       eprint = {1602.03894},
 primaryClass = {astro-ph.GA},
       adsurl = {https://ui.adsabs.harvard.edu/abs/2016ApJ...831....7S}
}

@ARTICLE{Dong2008,
       author = {{Dong}, Xiaobo and {Wang}, Tinggui and {Wang}, Jianguo and {Yuan}, Weimin and {Zhou}, Hongyan and {Dai}, Haifeng and {Zhang}, Kai},
        title = "{Broad-line Balmer decrements in blue active galactic nuclei}",
      journal = {\mnras},
         year = 2008,
        month = jan,
       volume = {383},
       number = {2},
        pages = {581-592},
          doi = {10.1111/j.1365-2966.2007.12560.x},
archivePrefix = {arXiv},
       eprint = {0710.1458},
 primaryClass = {astro-ph},
       adsurl = {https://ui.adsabs.harvard.edu/abs/2008MNRAS.383..581D}
}

@ARTICLE{Barrows2025,
       author = {{Barrows}, R. Scott and {Comerford}, Julia M. and {Negus}, James and {Muller-Sanchez}, Francisco},
        title = "{Recoiling Black Hole Candidates from Spatially Offset Broad Emission Lines in MaNGA}",
      journal = {\apj},
         year = 2025,
        month = oct,
       volume = {992},
       number = {1},
          eid = {38},
        pages = {38},
          doi = {10.3847/1538-4357/adfbf3},
       adsurl = {https://ui.adsabs.harvard.edu/abs/2025ApJ...992...38B}
}

@ARTICLE{Jadhav2021,
       author = {{Jadhav}, Yashashree and {Robinson}, Andrew and {Almeyda}, Triana and {Curran}, Rachel and {Marconi}, Alessandro},
        title = "{The spatially offset quasar E1821+643: new evidence for gravitational recoil}",
      journal = {\mnras},
         year = 2021,
        month = oct,
       volume = {507},
       number = {1},
        pages = {484-495},
          doi = {10.1093/mnras/stab2176},
archivePrefix = {arXiv},
       eprint = {2107.14711},
 primaryClass = {astro-ph.GA},
       adsurl = {https://ui.adsabs.harvard.edu/abs/2021MNRAS.507..484J}
}

@ARTICLE{Robinson2010,
       author = {{Robinson}, Andrew and {Young}, Stuart and {Axon}, David J. and {Kharb}, Preeti and {Smith}, James E.},
        title = "{Spectropolarimetric Evidence for a Kicked Supermassive Black Hole in the Quasar E1821+643}",
      journal = {\apjl},
         year = 2010,
        month = jul,
       volume = {717},
       number = {2},
        pages = {L122-L126},
          doi = {10.1088/2041-8205/717/2/L122},
archivePrefix = {arXiv},
       eprint = {1006.0993},
 primaryClass = {astro-ph.CO},
       adsurl = {https://ui.adsabs.harvard.edu/abs/2010ApJ...717L.122R}
}

@ARTICLE{Chiaberge2017,
       author = {{Chiaberge}, M. and {Ely}, J.~C. and {Meyer}, E.~T. and {Georganopoulos}, M. and {Marinucci}, A. and {Bianchi}, S. and {Tremblay}, G.~R. and {Hilbert}, B. and {Kotyla}, J.~P. and {Capetti}, A. and {Baum}, S.~A. and {Macchetto}, F.~D. and {Miley}, G. and {O'Dea}, C.~P. and {Perlman}, E.~S. and {Sparks}, W.~B. and {Norman}, C.},
        title = "{The puzzling case of the radio-loud QSO 3C 186: a gravitational wave recoiling black hole in a young radio source?}",
      journal = {\aap},
         year = 2017,
        month = apr,
       volume = {600},
          eid = {A57},
        pages = {A57},
          doi = {10.1051/0004-6361/201629522},
archivePrefix = {arXiv},
       eprint = {1611.05501},
 primaryClass = {astro-ph.GA},
       adsurl = {https://ui.adsabs.harvard.edu/abs/2017A&A...600A..57C}
}

@ARTICLE{Wu_Shen2022,
       author = {{Wu}, Qiaoya and {Shen}, Yue},
        title = "{A Catalog of Quasar Properties from Sloan Digital Sky Survey Data Release 16}",
      journal = {\apjs},
         year = 2022,
        month = dec,
       volume = {263},
       number = {2},
          eid = {42},
        pages = {42},
          doi = {10.3847/1538-4365/ac9ead},
archivePrefix = {arXiv},
       eprint = {2209.03987},
 primaryClass = {astro-ph.GA},
       adsurl = {https://ui.adsabs.harvard.edu/abs/2022ApJS..263...42W}
}

@ARTICLE{nanograv_15yr_astro,
       author = {{Agazie}, Gabriella and {Anumarlapudi}, Akash and {Archibald}, Anne M. and {Baker}, Paul T. and {B{\'e}csy}, Bence and {Blecha}, Laura and {Bonilla}, Alexander and {Brazier}, Adam and {Brook}, Paul R. and {Burke-Spolaor}, Sarah and {Burnette}, Rand and {Case}, Robin and {Casey-Clyde}, J. Andrew and {Charisi}, Maria and {Chatterjee}, Shami and {Chatziioannou}, Katerina and {Cheeseboro}, Belinda D. and {Chen}, Siyuan and {Cohen}, Tyler and {Cordes}, James M. and {Cornish}, Neil J. and {Crawford}, Fronefield and {Cromartie}, H. Thankful and {Crowter}, Kathryn and {Cutler}, Curt J. and {D'Orazio}, Daniel J. and {Decesar}, Megan E. and {Degan}, Dallas and {Demorest}, Paul B. and {Deng}, Heling and {Dolch}, Timothy and {Drachler}, Brendan and {Ferrara}, Elizabeth C. and {Fiore}, William and {Fonseca}, Emmanuel and {Freedman}, Gabriel E. and {Gardiner}, Emiko and {Garver-Daniels}, Nate and {Gentile}, Peter A. and {Gersbach}, Kyle A. and {Glaser}, Joseph and {Good}, Deborah C. and {G{\"u}ltekin}, Kayhan and {Hazboun}, Jeffrey S. and {Hourihane}, Sophie and {Islo}, Kristina and {Jennings}, Ross J. and {Johnson}, Aaron and {Jones}, Megan L. and {Kaiser}, Andrew R. and {Kaplan}, David L. and {Kelley}, Luke Zoltan and {Kerr}, Matthew and {Key}, Joey S. and {Laal}, Nima and {Lam}, Michael T. and {Lamb}, William G. and {Lazio}, T. Joseph W. and {Lewandowska}, Natalia and {Littenberg}, Tyson B. and {Liu}, Tingting and {Luo}, Jing and {Lynch}, Ryan S. and {Ma}, Chung-Pei and {Madison}, Dustin R. and {McEwen}, Alexander and {McKee}, James W. and {McLaughlin}, Maura A. and {McMann}, Natasha and {Meyers}, Bradley W. and {Meyers}, Patrick M. and {Mingarelli}, Chiara M.~F. and {Mitridate}, Andrea and {Natarajan}, Priyamvada and {Ng}, Cherry and {Nice}, David J. and {Ocker}, Stella Koch and {Olum}, Ken D. and {Pennucci}, Timothy T. and {Perera}, Benetge B.~P. and {Petrov}, Polina and {Pol}, Nihan S. and {Radovan}, Henri A. and {Ransom}, Scott M. and {Ray}, Paul S. and {Romano}, Joseph D. and {Runnoe}, Jessie C. and {Sardesai}, Shashwat C. and {Schmiedekamp}, Ann and {Schmiedekamp}, Carl and {Schmitz}, Kai and {Schult}, Levi and {Shapiro-Albert}, Brent J. and {Siemens}, Xavier and {Simon}, Joseph and {Siwek}, Magdalena S. and {Stairs}, Ingrid H. and {Stinebring}, Daniel R. and {Stovall}, Kevin and {Sun}, Jerry P. and {Susobhanan}, Abhimanyu and {Swiggum}, Joseph K. and {Taylor}, Jacob and {Taylor}, Stephen R. and {Turner}, Jacob E. and {Unal}, Caner and {Vallisneri}, Michele and {Vigeland}, Sarah J. and {Wachter}, Jeremy M. and {Wahl}, Haley M. and {Wang}, Qiaohong and {Witt}, Caitlin A. and {Wright}, David and {Young}, Olivia and {Nanograv Collaboration}},
        title = "{The NANOGrav 15 yr Data Set: Constraints on Supermassive Black Hole Binaries from the Gravitational-wave Background}",
      journal = {\apjl},
         year = 2023,
        month = aug,
       volume = {952},
       number = {2},
          eid = {L37},
        pages = {L37},
          doi = {10.3847/2041-8213/ace18b},
archivePrefix = {arXiv},
       eprint = {2306.16220},
 primaryClass = {astro-ph.HE},
       adsurl = {https://ui.adsabs.harvard.edu/abs/2023ApJ...952L..37A}
}

@ARTICLE{PTF_candidates_Charisi_et_al2016,
       author = {{Charisi}, M. and {Bartos}, I. and {Haiman}, Z. and {Price-Whelan}, A.~M. and {Graham}, M.~J. and {Bellm}, E.~C. and {Laher}, R.~R. and {M{\'a}rka}, S.},
        title = "{A population of short-period variable quasars from PTF as supermassive black hole binary candidates}",
      journal = {\mnras},
         year = 2016,
        month = dec,
       volume = {463},
       number = {2},
        pages = {2145-2171},
          doi = {10.1093/mnras/stw1838},
archivePrefix = {arXiv},
       eprint = {1604.01020},
 primaryClass = {astro-ph.GA},
       adsurl = {https://ui.adsabs.harvard.edu/abs/2016MNRAS.463.2145C}
}

@ARTICLE{CRTS_candidates_Graham_et_al2015,
       author = {{Graham}, Matthew J. and {Djorgovski}, S.~G. and {Stern}, Daniel and {Drake}, Andrew J. and {Mahabal}, Ashish A. and {Donalek}, Ciro and {Glikman}, Eilat and {Larson}, Steve and {Christensen}, Eric},
        title = "{A systematic search for close supermassive black hole binaries in the Catalina Real-time Transient Survey}",
      journal = {\mnras},
         year = 2015,
        month = oct,
       volume = {453},
       number = {2},
        pages = {1562-1576},
          doi = {10.1093/mnras/stv1726},
archivePrefix = {arXiv},
       eprint = {1507.07603},
 primaryClass = {astro-ph.GA},
       adsurl = {https://ui.adsabs.harvard.edu/abs/2015MNRAS.453.1562G}
}

@ARTICLE{Sigma_dust_proxy_Ledoux_et_al2015,
       author = {{Ledoux}, C. and {Noterdaeme}, P. and {Petitjean}, P. and {Srianand}, R.},
        title = "{Neutral atomic-carbon quasar absorption-line systems at z> 1.5. Sample selection, H i content, reddening, and 2175 {\r{A}} extinction feature}",
      journal = {\aap},
         year = 2015,
        month = aug,
       volume = {580},
          eid = {A8},
        pages = {A8},
          doi = {10.1051/0004-6361/201424122},
archivePrefix = {arXiv},
       eprint = {1504.07254},
 primaryClass = {astro-ph.GA},
       adsurl = {https://ui.adsabs.harvard.edu/abs/2015A&A...580A...8L}
}

@ARTICLE{SDSS-DR16Q,
       author = {{Lyke}, Brad W. and {Higley}, Alexandra N. and {McLane}, J.~N. and {Schurhammer}, Danielle P. and {Myers}, Adam D. and {Ross}, Ashley J. and {Dawson}, Kyle and {Chabanier}, Sol{\`e}ne and {Martini}, Paul and {Busca}, Nicol{\'a}s G. and {Mas des Bourboux}, H{\'e}lion du and {Salvato}, Mara and {Streblyanska}, Alina and {Zarrouk}, Pauline and {Burtin}, Etienne and {Anderson}, Scott F. and {Bautista}, Julian and {Bizyaev}, Dmitry and {Brandt}, W.~N. and {Brinkmann}, Jonathan and {Brownstein}, Joel R. and {Comparat}, Johan and {Green}, Paul and {de la Macorra}, Axel and {Mu{\~n}oz Guti{\'e}rrez}, Andrea and {Hou}, Jiamin and {Newman}, Jeffrey A. and {Palanque-Delabrouille}, Nathalie and {P{\^a}ris}, Isabelle and {Percival}, Will J. and {Petitjean}, Patrick and {Rich}, James and {Rossi}, Graziano and {Schneider}, Donald P. and {Smith}, Alexander and {Vivek}, M. and {Weaver}, Benjamin Alan},
        title = "{The Sloan Digital Sky Survey Quasar Catalog: Sixteenth Data Release}",
      journal = {\apjs},
         year = 2020,
        month = sep,
       volume = {250},
       number = {1},
          eid = {8},
        pages = {8},
          doi = {10.3847/1538-4365/aba623},
archivePrefix = {arXiv},
       eprint = {2007.09001},
 primaryClass = {astro-ph.GA},
       adsurl = {https://ui.adsabs.harvard.edu/abs/2020ApJS..250....8L}
}

@ARTICLE{SDSS-DR12Q,
       author = {{P{\^a}ris}, Isabelle and {Petitjean}, Patrick and {Ross}, Nicholas P. and {Myers}, Adam D. and {Aubourg}, {\'E}ric and {Streblyanska}, Alina and {Bailey}, Stephen and {Armengaud}, {\'E}ric and {Palanque-Delabrouille}, Nathalie and {Y{\`e}che}, Christophe and {Hamann}, Fred and {Strauss}, Michael A. and {Albareti}, Franco D. and {Bovy}, Jo and {Bizyaev}, Dmitry and {Niel Brandt}, W. and {Brusa}, Marcella and {Buchner}, Johannes and {Comparat}, Johan and {Croft}, Rupert A.~C. and {Dwelly}, Tom and {Fan}, Xiaohui and {Font-Ribera}, Andreu and {Ge}, Jian and {Georgakakis}, Antonis and {Hall}, Patrick B. and {Jiang}, Linhua and {Kinemuchi}, Karen and {Malanushenko}, Elena and {Malanushenko}, Viktor and {McMahon}, Richard G. and {Menzel}, Marie-Luise and {Merloni}, Andrea and {Nandra}, Kirpal and {Noterdaeme}, Pasquier and {Oravetz}, Daniel and {Pan}, Kaike and {Pieri}, Matthew M. and {Prada}, Francisco and {Salvato}, Mara and {Schlegel}, David J. and {Schneider}, Donald P. and {Simmons}, Audrey and {Viel}, Matteo and {Weinberg}, David H. and {Zhu}, Liu},
        title = "{The Sloan Digital Sky Survey Quasar Catalog: Twelfth data release}",
      journal = {\aap},
         year = 2017,
        month = jan,
       volume = {597},
          eid = {A79},
        pages = {A79},
          doi = {10.1051/0004-6361/201527999},
archivePrefix = {arXiv},
       eprint = {1608.06483},
 primaryClass = {astro-ph.GA},
       adsurl = {https://ui.adsabs.harvard.edu/abs/2017A&A...597A..79P}
}

@ARTICLE{offset_detectability_Kelley2021,
       author = {{Kelley}, Luke Zoltan},
        title = "{Basic considerations for the observability of kinematically offset binary AGN}",
      journal = {\mnras},
         year = 2021,
        month = jan,
       volume = {500},
       number = {3},
        pages = {4065-4077},
          doi = {10.1093/mnras/staa3219},
archivePrefix = {arXiv},
       eprint = {2005.10255},
 primaryClass = {astro-ph.HE},
       adsurl = {https://ui.adsabs.harvard.edu/abs/2021MNRAS.500.4065K}
}

@ARTICLE{RecoilingBH_Blecha_et_al2016,
       author = {{Blecha}, Laura and {Sijacki}, Debora and {Kelley}, Luke Zoltan and {Torrey}, Paul and {Vogelsberger}, Mark and {Nelson}, Dylan and {Springel}, Volker and {Snyder}, Gregory and {Hernquist}, Lars},
        title = "{Recoiling black holes: prospects for detection and implications of spin alignment}",
      journal = {\mnras},
         year = 2016,
        month = feb,
       volume = {456},
       number = {1},
        pages = {961-989},
          doi = {10.1093/mnras/stv2646},
archivePrefix = {arXiv},
       eprint = {1508.01524},
 primaryClass = {astro-ph.GA},
       adsurl = {https://ui.adsabs.harvard.edu/abs/2016MNRAS.456..961B}
}

@ARTICLE{RecoilingBH_Komossa2012,
       author = {{Komossa}, S.},
        title = "{Recoiling Black Holes: Electromagnetic Signatures, Candidates, and Astrophysical Implications}",
      journal = {Advances in Astronomy},
         year = 2012,
        month = jan,
       volume = {2012},
          eid = {364973},
        pages = {364973},
          doi = {10.1155/2012/364973},
archivePrefix = {arXiv},
       eprint = {1202.1977},
 primaryClass = {astro-ph.CO},
       adsurl = {https://ui.adsabs.harvard.edu/abs/2012AdAst2012E..14K}
}

@ARTICLE{RecoilingBH_Blecha_et_al2011,
       author = {{Blecha}, Laura and {Cox}, Thomas J. and {Loeb}, Abraham and {Hernquist}, Lars},
        title = "{Recoiling black holes in merging galaxies: relationship to active galactic nucleus lifetimes, starbursts and the M$_{BH}$-{\ensuremath{\sigma}}$_{*}$ relation}",
      journal = {\mnras},
         year = 2011,
        month = apr,
       volume = {412},
       number = {4},
        pages = {2154-2182},
          doi = {10.1111/j.1365-2966.2010.18042.x},
archivePrefix = {arXiv},
       eprint = {1009.4940},
 primaryClass = {astro-ph.CO},
       adsurl = {https://ui.adsabs.harvard.edu/abs/2011MNRAS.412.2154B}
}

@ARTICLE{RecoilingBH_Bonning_et_al2007,
       author = {{Bonning}, E.~W. and {Shields}, G.~A. and {Salviander}, S.},
        title = "{Recoiling Black Holes in Quasars}",
      journal = {\apjl},
         year = 2007,
        month = sep,
       volume = {666},
       number = {1},
        pages = {L13-L16},
          doi = {10.1086/521674},
archivePrefix = {arXiv},
       eprint = {0705.4263},
 primaryClass = {astro-ph},
       adsurl = {https://ui.adsabs.harvard.edu/abs/2007ApJ...666L..13B}
}

@ARTICLE{EjectedQSO_timescale_Loeb2007,
       author = {{Loeb}, Abraham},
        title = "{Observable Signatures of a Black Hole Ejected by Gravitational-Radiation Recoil in a Galaxy Merger}",
      journal = {\prl},
         year = 2007,
        month = jul,
       volume = {99},
       number = {4},
          eid = {041103},
        pages = {041103},
          doi = {10.1103/PhysRevLett.99.041103},
archivePrefix = {arXiv},
       eprint = {astro-ph/0703722},
 primaryClass = {astro-ph},
       adsurl = {https://ui.adsabs.harvard.edu/abs/2007PhRvL..99d1103L}
}

@ARTICLE{Tanaka_Haiman2009,
       author = {{Tanaka}, Takamitsu and {Haiman}, Zolt{\'a}n},
        title = "{The Assembly of Supermassive Black Holes at High Redshifts}",
      journal = {\apj},
         year = 2009,
        month = may,
       volume = {696},
       number = {2},
        pages = {1798-1822},
          doi = {10.1088/0004-637X/696/2/1798},
archivePrefix = {arXiv},
       eprint = {0807.4702},
 primaryClass = {astro-ph},
       adsurl = {https://ui.adsabs.harvard.edu/abs/2009ApJ...696.1798T}
}

@ARTICLE{oscill_Momossa_Merritt2008,
       author = {{Komossa}, S. and {Merritt}, David},
        title = "{Gravitational Wave Recoil Oscillations of Black Holes: Implications for Unified Models of Active Galactic Nuclei}",
      journal = {\apjl},
         year = 2008,
        month = dec,
       volume = {689},
       number = {2},
        pages = {L89},
          doi = {10.1086/595883},
archivePrefix = {arXiv},
       eprint = {0811.1037},
 primaryClass = {astro-ph},
       adsurl = {https://ui.adsabs.harvard.edu/abs/2008ApJ...689L..89K}
}

@ARTICLE{recoil_Madau_Quataert2004,
       author = {{Madau}, Piero and {Quataert}, Eliot},
        title = "{The Effect of Gravitational-Wave Recoil on the Demography of Massive Black Holes}",
      journal = {\apjl},
         year = 2004,
        month = may,
       volume = {606},
       number = {1},
        pages = {L17-L20},
          doi = {10.1086/421017},
archivePrefix = {arXiv},
       eprint = {astro-ph/0403295},
 primaryClass = {astro-ph},
       adsurl = {https://ui.adsabs.harvard.edu/abs/2004ApJ...606L..17M}
}

@ARTICLE{black_hole_recoil_Healy2014,
       author = {{Healy}, James and {Lousto}, Carlos O. and {Zlochower}, Yosef},
        title = "{Remnant mass, spin, and recoil from spin aligned black-hole binaries}",
      journal = {\prd},
         year = 2014,
        month = nov,
       volume = {90},
       number = {10},
          eid = {104004},
        pages = {104004},
          doi = {10.1103/PhysRevD.90.104004},
archivePrefix = {arXiv},
       eprint = {1406.7295},
 primaryClass = {gr-qc},
       adsurl = {https://ui.adsabs.harvard.edu/abs/2014PhRvD..90j4004H}
}

@ARTICLE{Sesana_at_al2005,
       author = {{Sesana}, Alberto and {Haardt}, Francesco and {Madau}, Piero and {Volonteri}, Marta},
        title = "{The Gravitational Wave Signal from Massive Black Hole Binaries and Its Contribution to the LISA Data Stream}",
      journal = {\apj},
         year = 2005,
        month = apr,
       volume = {623},
       number = {1},
        pages = {23-30},
          doi = {10.1086/428492},
archivePrefix = {arXiv},
       eprint = {astro-ph/0409255},
 primaryClass = {astro-ph},
       adsurl = {https://ui.adsabs.harvard.edu/abs/2005ApJ...623...23S}
}

@ARTICLE{Volonteri_Haardt_Madau2003,
       author = {{Volonteri}, Marta and {Haardt}, Francesco and {Madau}, Piero},
        title = "{The Assembly and Merging History of Supermassive Black Holes in Hierarchical Models of Galaxy Formation}",
      journal = {\apj},
         year = 2003,
        month = jan,
       volume = {582},
       number = {2},
        pages = {559-573},
          doi = {10.1086/344675},
archivePrefix = {arXiv},
       eprint = {astro-ph/0207276},
 primaryClass = {astro-ph},
       adsurl = {https://ui.adsabs.harvard.edu/abs/2003ApJ...582..559V}
}

@ARTICLE{Milosavljevi_Merritt2001,
       author = {{Milosavljevi{\'c}}, Milo{\v{s}} and {Merritt}, David},
        title = "{Formation of Galactic Nuclei}",
      journal = {\apj},
         year = 2001,
        month = dec,
       volume = {563},
       number = {1},
        pages = {34-62},
          doi = {10.1086/323830},
archivePrefix = {arXiv},
       eprint = {astro-ph/0103350},
 primaryClass = {astro-ph},
       adsurl = {https://ui.adsabs.harvard.edu/abs/2001ApJ...563...34M}
}

@ARTICLE{Begelman_Blandford_Rees1980,
       author = {{Begelman}, M.~C. and {Blandford}, R.~D. and {Rees}, M.~J.},
        title = "{Massive black hole binaries in active galactic nuclei}",
      journal = {\nat},
         year = 1980,
        month = sep,
       volume = {287},
       number = {5780},
        pages = {307-309},
          doi = {10.1038/287307a0},
       adsurl = {https://ui.adsabs.harvard.edu/abs/1980Natur.287..307B}
}

@ARTICLE{Comeford_et_al2009,
       author = {{Comerford}, Julia M. and {Gerke}, Brian F. and {Newman}, Jeffrey A. and {Davis}, Marc and {Yan}, Renbin and {Cooper}, Michael C. and {Faber}, S.~M. and {Koo}, David C. and {Coil}, Alison L. and {Rosario}, D.~J. and {Dutton}, Aaron A.},
        title = "{Inspiralling Supermassive Black Holes: A New Signpost for Galaxy Mergers}",
      journal = {\apj},
         year = 2009,
        month = jun,
       volume = {698},
       number = {1},
        pages = {956-965},
          doi = {10.1088/0004-637X/698/1/956},
archivePrefix = {arXiv},
       eprint = {0810.3235},
 primaryClass = {astro-ph},
       adsurl = {https://ui.adsabs.harvard.edu/abs/2009ApJ...698..956C}
}

@ARTICLE{Lacey_Cole1993,
       author = {{Lacey}, Cedric and {Cole}, Shaun},
        title = "{Merger rates in hierarchical models of galaxy formation}",
      journal = {\mnras},
         year = 1993,
        month = jun,
       volume = {262},
       number = {3},
        pages = {627-649},
          doi = {10.1093/mnras/262.3.627},
       adsurl = {https://ui.adsabs.harvard.edu/abs/1993MNRAS.262..627L}
}

@ARTICLE{Richstone1998,
       author = {{Richstone}, D. and {Ajhar}, E.~A. and {Bender}, R. and {Bower}, G. and {Dressler}, A. and {Faber}, S.~M. and {Filippenko}, A.~V. and {Gebhardt}, K. and {Green}, R. and {Ho}, L.~C. and {Kormendy}, J. and {Lauer}, T.~R. and {Magorrian}, J. and {Tremaine}, S.},
        title = "{Supermassive black holes and the evolution of galaxies.}",
      journal = {\nat},
         year = 1998,
        month = oct,
       volume = {385},
       number = {6701},
        pages = {A14},
          doi = {10.48550/arXiv.astro-ph/9810378},
archivePrefix = {arXiv},
       eprint = {astro-ph/9810378},
 primaryClass = {astro-ph},
       adsurl = {https://ui.adsabs.harvard.edu/abs/1998Natur.395A..14R}
}

@ARTICLE{AGNOsc_proc2017,
       author = {{Raffai}, P. and {B{\'e}csy}, B. and {Haiman}, Z. and {Frei}, Z.},
        title = "{A Statistical Method for Detecting Gravitational Recoils of Supermassive Black Holes in Active Galactic Nuclei}",
    journal = {Proceedings IAU Symposium},
         year = 2017,
       editor = {{Gomboc}, Andreja},
       volume = {324},
        month = jan,
        pages = {227-230},
          doi = {10.1017/S1743921317000734},
       adsurl = {https://ui.adsabs.harvard.edu/abs/2017IAUS..324..227R}
}

@ARTICLE{AGNOsc_theory2016,
       author = {{Raffai}, P. and {Haiman}, Z. and {Frei}, Z.},
        title = "{A statistical method to search for recoiling supermassive black holes in active galactic nuclei}",
      journal = {\mnras},
         year = 2016,
        month = jan,
       volume = {455},
       number = {1},
        pages = {484-492},
          doi = {10.1093/mnras/stv2371},
archivePrefix = {arXiv},
       eprint = {1509.02075},
 primaryClass = {astro-ph.GA},
       adsurl = {https://ui.adsabs.harvard.edu/abs/2016MNRAS.455..484R}
}




\bsp	
\label{lastpage}
\end{document}